\documentclass[journal,ruled,vlined,linesnumbered]{IEEEtran}
\usepackage[T1]{fontenc}
\usepackage[utf8]{inputenc}
\usepackage{algorithm2e}
\usepackage{amsmath}
\usepackage{amssymb}
\usepackage{graphicx}
\usepackage[bookmarks=true,bookmarksnumbered=true,bookmarksopen=true,bookmarksopenlevel=1,
 breaklinks=false,pdfborder={0 0 1},backref=false,colorlinks=false]
 {hyperref}
\hypersetup{pdftitle={Your~Title},
 pdfauthor={Your Name},
 pdfborderstyle=,pdfpagelayout=OneColumn,pdfnewwindow=true,pdfstartview=XYZ,plainpages=false}



\begin{document}
\title{Learning to Beamform for Cooperative\\
 Localization and Communication: \\
 A Link Heterogeneous GNN-Based Approach}
\author{\IEEEauthorblockN{Lixiang Lian, \IEEEmembership{Member, IEEE}, Chuanqi
Bai, Yihan Xu, Huanyu Dong, Rui Cheng and Shunqing Zhang, \IEEEmembership{Senior Member, IEEE}}
\thanks{Lixiang Lian, Chuanqi Bai, Yihan Xu and Huanyu Dong are with the School
of Information Science and Technology, ShanghaiTech University, Shanghai
201210, China (e-mail: lianlx@shanghaitech.edu.cn; baichq2024@shanghaitech.edu.cn;
xuyh3@shanghaitech.edu.cn; huanyu.dong0425@gmail.com).} \thanks{Rui Cheng and Shunqing Zhang are with the Shanghai Institute for Advanced
Communication and Data Science, Key Laboratory of Specialty Fiber
Optics and Optical Access Networks, School of Information and Communication
Engineering, Shanghai University, Shanghai, 200444, China (e-mail:
cora\_cheng@shu.edu.cn; shunqing@shu.edu.cn).}}
\maketitle
\begin{abstract}
Integrated sensing and communication (ISAC) has emerged as a key enabler
for next-generation wireless networks, supporting advanced applications
such as high-precision localization and environment reconstruction.
Cooperative ISAC (CoISAC) further enhances these capabilities by enabling
multiple base stations (BSs) to jointly optimize communication and
sensing performance through coordination. However, CoISAC beamforming
design faces significant challenges due to system heterogeneity, large-scale
problem complexity, and sensitivity to parameter estimation errors.
Traditional deep learning-based techniques fail to exploit the unique
structural characteristics of CoISAC systems, thereby limiting their
ability to enhance system performance. To address these challenges,
we propose a Link-Heterogeneous Graph Neural Network (LHGNN) for joint
beamforming in CoISAC systems. Unlike conventional approaches, LHGNN
models communication and sensing links as heterogeneous nodes and
their interactions as edges, enabling the capture of the heterogeneous
nature and intricate interactions of CoISAC systems. Furthermore,
a graph attention mechanism is incorporated to dynamically adjust
node and link importance, improving robustness to channel and position
estimation errors. Numerical results demonstrate that the proposed
attention-enhanced LHGNN achieves superior communication rates while
maintaining sensing accuracy under power constraints. The proposed
method also exhibits strong robustness to communication channel and
position estimation error. 
\end{abstract}

\begin{IEEEkeywords}
Cooperative ISAC, direct localization, beamforming, heterogeneous
graph neural networks. 
\end{IEEEkeywords}

\IEEEpeerreviewmaketitle{\thispagestyle{empty}}

\section{Introduction}

\label{sec:intro}

\IEEEPARstart{I}{ntegrated} sensing and communication (ISAC) has
become a key technology in next-generation wireless communication
networks. By integrating the sensing and communication functionalities,
the ISAC system can support various cutting-edge applications such
as high-precision localization, environment reconstruction, smart
healthcare, etc\cite{tan2021integrated}. Utilizing the same spectrum
and hardware resources, one of the key issues of ISAC is to guarantee
both sensing and communication performance. Cooperative ISAC (CoISAC)
has drawn much attention due to its potential to enhance both communication
and sensing services. By connecting to a common central processing
unit (CPU), multiple base stations (BSs) cooperatively provide joint
communication and sensing services. Coordination among BSs can better
manage the interferences among communication users to improve the
communication data rate. Additionally, multiple BSs provide diverse
spatial information, enhancing the precision of sensing tasks. By
combining information from various sources relying on the widely-used
cloud radio access network (C-RAN) architecture \cite{wu2015cloudRAN},
CoISAC enhances efficiency and robustness for both communication and
sensing tasks\cite{liu2020cloud}.

Despite the potential benefits of CoISAC, the design of a joint beamforming
scheme for multi-BS CoISAC systems to simultaneously optimize multi-user
communication performance and target sensing performance presents
significant challenges. First, CoISAC system exhibits significant
heterogeneity across mutliple dimensions. Communication and sensing
functions differ in optimization objectives, link characteristics,
and signal responses, for communication users and sensing targets.
Moreover, when multiple BSs collaborate, heterogeneity arises among
BSs in terms of varying resource budgets, physical environments, and
channel conditions associated with communication and sensing users.
The system's multi-objective nature and complex link topology significantly
increase the complexity of the joint beamforming problem. Second,
the shift towards 6G technology results in an expanded problem scale
due to the increased number of antennas, higher network density, and
the cooperation among multiple BSs. This, in turn, leads to a significantly
larger-scale transmit beamforming design problem in CoISAC systems.
Third, accurate channel information in communication and precise target
information in sensing are essential for directing beamforming design.
Errors in estimation can lead to misdirected beams, thereby degrading
both communication and sensing performance \cite{liu2022survey}.
Therefore, designing beamforming schemes that are robust to estimation
errors in CoISAC systems is of great importance, which, however, can
induce additional complexities.

To address these difficulties, previous works mainly utilize traditional
optimization-based algorithms for beamforming design. \cite{chen2023comp_clustering}
investigated the joint waveform design and BS clustering problem in
CoISAC, considering sensing performance, communication performance,
and the backhaul capacity limitation. Fractional programming (FP)
techniques such as Dinkel-bach's transform and quadratic transform
are adopted to tackle the fractional signal-to-interference-plus-noise
ratio (SINR) objectives. Successive convex approximation (SCA) and
semidefinite relaxation (SDR) are utilized to handle the non-convexity
of the clustering and beamforming, respectively. \cite{babu2024precoding}
studied the minimization of the angular parameter Cramér-Rao bound
(CRB) in mono-static and bi-static CoISAC scenarios. FP and SDR are
also adopted to convexify the beamforming design problem. \cite{gao2023SPEB}
investigated CoISAC beamforming design considering squared position
error bound (SPEB) and communication rate requirements. SCA technique
is adopted to tackle the non-convexity of BS scheduling, fronthaul,
and SPEB constraints. However, these traditional methods have several
drawbacks. First, traditional algorithms handle the non-convexity
of the optimization problem using approximation methods, such as SCA,
FP, and SDR, which can only derive sub-optimal solutions. Second,
traditional algorithms usually need to iterate until they converge,
and also involve high-complexity operations such as matrix inversion
and decomposition. Thus, traditional algorithms usually have high
complexity and poor scalability, causing extremely high computational
latency. Third, although robust beamforming design has been considered
in ISAC systems, such as \cite{ISACrobust:yang2024coordinatedy,ISACrobust:zhang2024energy},
to handle information uncertainties, the resulting optimization algorithms
are extremely complex, and not well-suited for practical systems.
In addition to their complexities, these works rely on uncertainty
sets of parameters defined by specific thresholds. This inflexibility
can lead to either overly pessimistic or overly optimistic solutions.

Deep learning has attracted a lot of attention in physical layer optimization
due to its excellent approximation capability and low inference complexity.
A lot of works have investigated deep learning-based ISAC beamforming
design. \cite{ISACLSTM_liu2022learning} proposed a convolutional
long short-term memory (LSTM) network-based algorithm to extract the
features of historical channels and predict the ISAC beamforming in
the vehicle networks, which reduces the dependency on real-time CSI.
\cite{liu2022ppo_bf} proposed a deep reinforcement learning-based
algorithm to solve the radar beam pattern error-constrained sum-rate
maximization problem in an intelligent reflecting surface (IRS)-aided
ISAC scenario. \cite{yang2024deep} studied deep learning-based ISAC
beamforming algorithm, aimed at maximizing the average SINR for multi-user
communication while guaranteeing the detection probability of radar
targets. \cite{qi2024deep_uplink_ISAC} investigated the joint sensing
transmit waveform and communication receive beamforming design with
the objective of maximizing the weighted sum of the normalized sensing
rate and normalized communication rate in an uplink scenario leveraging
the deep learning techniques. However, current deep learning-based
ISAC systems primarily employ general-purpose neural networks, such
as convolutional neural networks (CNNs) \cite{ISACCNN_chen2024complex},
LSTMs\cite{ISACLSTM_liu2022learning}, or feedforward neural networks
(FNNs)\cite{ISACFNN_pulkkinen2024model}, for beamforming design,
without leveraging the specific network topology of ISAC systems.
In particular, considering the heterogeneous nature of the CoISAC
system's network topology and the optimization problem itself, directly
applying these general-purpose neural networks to CoISAC for beamforming
design faces several challenges\cite{SUV_shi2023largescale}. First,
the training efficiency is low, severely impacting learning performance.
Second, the scalability and generalizability of these networks are
poor, making them unsuitable for large-scale CoISAC topologies. Therefore,
developing specialized neural networks tailored to beamforming optimization
in CoISAC systems is essential.

Graph neural networks (GNN) can effectively incorporate graph structures,
modeling node attributes and relationships between nodes to explore
hidden features in graph-structured data \cite{shen2022graph}. Therefore,
GNNs have been introduced to solve a broad range of optimization problems
in wireless networks, such as multi-user beamforming problem in cell-free
networks \cite{GNNAPP_yan2024cellfree}, joint sensing and communication
service scheduling and vehicle association in vehicular networks \cite{GNNAPP_li2024gnn_isac},
joint Sub-6G and mmWave networks \cite{GNNAPP_huang2024sub}, etc.
By exploiting the inherent property of the problem, GNN can achieve
near-optimal performance with faster training speed, fewer training
samples, better scalability and generalizability than general-purpose
networks \cite{SUV_shi2023largescale}. However, to the best of our
knowledge, there are no previous works focused on CoISAC beamforming
optimization based on GNN. Due to the heterogeneity in CoISAC systems,
traditional bipartite graph modeling cannot adequately represent the
complex link structures and multi-objective nature of the network.
To effectively model the heterogeneous link types and their complex
interaction in CoISAC systems, a link-heterogeneous GNN (LHGNN) will
be proposed in this paper. Instead of moding each device as a heterogenous
node, we model the different types of links, such as communication
and sensing links that connect multiple ISAC-enabled BSs with multiple
communication users and sensing target, as heterogeneous nodes. This
approach allows us to construct a link-node-based graph structure,
which fully capture the heterogeneous nature and intricate interactions
of CoISAC systems. To improve the robustness of the LHGNN to parameter
estimation errors, we further incorporate a graph attention mechanism
into LHGNN. This enables the GNN to learn the optimal weights during
the information aggregation, adapting to the input parameter errors
and thereby reducing the impact of parameter uncertainties on the
final beamforming output. 

In this paper, we focus on the scenario of multi-BS collaborative
localization and communication, innovatively introducing LHGNN for
beamforming design in CoISAC systems. Graph attention is adopted to
improve the robustness to channel and target position estimation errors.
Our main contributions are summarized as follows: 
\begin{itemize}
\item \textbf{Link-Heterogeneous GNN Network for Joint Beamforming in CoISAC
Systems:} We consider multi-BSs are collaboratively supporting multi-user
communication and target direct localization simultaneously. We derive
the squared position error bound (SPEB) performance metric in CoISAC
systems to characterize the direct localization error. Then an SPEB
and power constrained sum-rate maximization problem is formulated
to jointly design the communication and sensing beamforming vectors
at multiple BSs. To fully capture the heterogeneity of CoISAC networks,
we model the different types of links as heterogeneous nodes and interferences
among links as edges to construct the graph. Compared to the traditional
bipartite graph model, the proposed LHGNN effectively models the heterogeneous
structure of communication and sensing links and fully exploiting
the dependencies between different nodes. 
\item \textbf{Graph Attention Mechanism-Enhanced LHGNN for Improved Robustness}:
We introduce the graph attention mechanism into the LHGNN to enhance
its robustness to errors in communication channels and position estimation.
This mechanism allows the model to dynamically adjust the importance
of different nodes and links, effectively mitigating the impact of
inaccurate input information on the final beamforming outputs and
improving the robustness of CoISAC systems in the presence of parameter
uncertainties. 
\item Numerical results have demonstrated that the proposed attention-enhanced
LHGNN achieves better performance in terms of achievable communication
rate while ensuring sensing performance and power consumption. Additionally,
the proposed algorithm has exhibited good robustness to channel and
position estimation errors. 
\end{itemize}
The rest of paper is organized as follows. In Section \ref{sec:sys_model},
we present the signal models of CoISAC systems, derive the squared
position error bound for cooperative localization, and formulate the
joint beamforming optimization problem. In Section \ref{sec:algo_trad},
we propose the framework of LHGNN, as well as the graph attention
mechanism to effectively solve the joint beamforming problem in CoISAC
systems. Finally, the extensive simulation results are provided in
Section \ref{sec:Simulation-Results} and conclusions are drawn in
Section \ref{sec:Conclusions}.

\section{System Model}

\label{sec:sys_model}

\begin{figure}[htbp]
\begin{centering}
\includegraphics[width=0.35\textwidth]{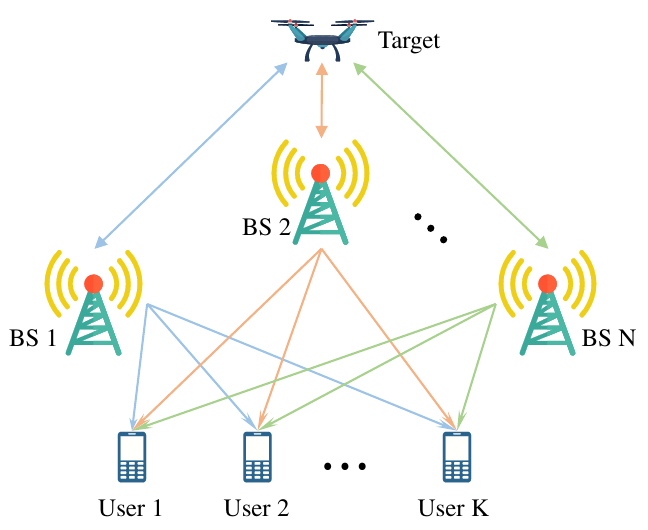} 
\par\end{centering}
\caption{Illustration of the proposed cooperative ISAC system.}
\label{fig:system_model} 
\end{figure}

In this section, we investigate a CoISAC system involving multiple
dual-functional BSs and multiple users, which is illustrated in Fig.
\ref{fig:system_model}. In particular, $N$ BSs cooperatively serve
$K$ single-antenna downlink users, while estimating the position
of a single sensing target\footnote{Although a single target is considered during the problem formulation,
the proposed algorithm can be readily extended to multiple sensing
targets scenario by including multiple sensing links.}. Each BS has a uniform planar array (UPA) with $L$ antenna elements.
The BSs are connected to a central processor (CP) through ideal backhaul
links. The transmit data assignment and precoding matrices design
are conducted at the CP and then distributed to each BS via fronthaul
links. $\mathcal{N}=\{1,...,N\}$ and $\mathcal{K}=\{1,...,K\}$ denote
the indices of BSs and users, respectively. 

\subsection{Communication Model}

\label{sec:com_model} Each BS transmits symbols $\mathbf{s}=[s^{s},s_{1}^{c},...,s_{K}^{c}]^{T}$,
where $s_{k}^{c}$ denotes the symbol for the $k$-th downlink user
from BS and $s^{s}$ is the dedicated symbol for target sensing. We
consider all symbols to be unit-power and independent from each other,
i.e., $\mathbb{E}[{\mathbf{s}}{\mathbf{s}}^{H}]=\mathbf{I}_{K+1}$.
The transmit signal of the $n$-th BS can be formulated as
\begin{equation}
\boldsymbol{x}_{n}=\mathbf{P}_{n}\mathbf{s}=\boldsymbol{p}_{n}^{s}s^{s}+\sum_{k=1}^{K}\boldsymbol{p}_{n,k}^{c}s_{k}^{c},\label{eq::transmit_signal}
\end{equation}
where $\mathbf{P}_{n}=[\boldsymbol{p}_{n}^{s},\boldsymbol{p}_{n,1}^{c},\ldots,\boldsymbol{p}_{n,K}^{c}]\in\mathbb{C}^{L\times(K+1)}$.
Denote $\boldsymbol{p}^{s}=[\boldsymbol{p}_{1}^{s,T},\ldots,\boldsymbol{p}_{N}^{s,T}]^{T}\in\mathbb{C}^{NL\times1}$
as the coordinated sensing beamforming vectors of $N$ BSs, $\boldsymbol{p}_{k}^{c}=\left[\boldsymbol{p}_{1,k}^{c,T},...,\boldsymbol{p}_{N,k}^{c,T}\right]^{T}\in\mathbb{C}^{NL\times1}$
as the coordinated communication beamforming vectors of $N$ BSs for
the $k$-th user, and $\boldsymbol{x}=[\boldsymbol{x}_{1}^{T},...,...,\boldsymbol{x}_{N}^{T}]^{T}\in\mathbb{C}^{NL\times1}$
as the concatenated transmit signal of $N$ BSs, respectively. Then
the received signal of the $k$-th user is formulated as
\begin{align}
y_{k} & =\boldsymbol{h}_{k}^{H}\boldsymbol{x}+n_{k}\nonumber \\
 & =\boldsymbol{h}_{k}^{H}\boldsymbol{p}_{k}^{c}s_{k}^{c}+\left(\boldsymbol{h}_{k}^{H}\boldsymbol{p}^{s}s^{s}+\sum_{j\in\mathcal{K},j\ne k}\boldsymbol{h}_{k}^{H}\boldsymbol{p}_{j}^{c}s_{j}^{c}\right)+n_{k},\label{eq::signal_com}
\end{align}
where $\boldsymbol{h}_{k}=[\boldsymbol{h}_{k,1}^{H},\boldsymbol{h}_{k,2}^{H},\ldots,\boldsymbol{h}_{k,N}^{H}]^{H}\in\mathbb{C}^{NL\times1}$
is the concatenated channel vector between $N$ BSs and the $k$-th
user. Without loss of generality, we consider $n_{k}\sim\mathcal{CN}(0,\sigma_{}^{2})$
is the complex additive white Gaussian noise with zero mean and variance
$\sigma^{2}$. Then the communication SINR can be expressed as
\begin{align}
{\rm SINR}_{k}=\frac{\left|\boldsymbol{h}_{k}^{H}\boldsymbol{p}_{k}^{c}\right|^{2}}{\left|\boldsymbol{h}_{k}^{H}\boldsymbol{p}^{s}\right|^{2}+\sum_{j\in\mathcal{K},j\ne k}\left|\boldsymbol{h}_{k}^{H}\boldsymbol{p}_{j}^{c}\right|^{2}+\sigma^{2}}.\label{eq::SINR}
\end{align}
According to \eqref{eq::SINR}, the achievable communication rate
of user $k$ is given by
\begin{align}
R_{k}=\log_{2}\left(1+{\rm SINR}_{k}\right).\label{eq:rate}
\end{align}

\subsection{Sensing Model}

\label{sec:sen_model}

\begin{figure}[htbp]
\begin{centering}
\includegraphics[width=0.35\textwidth]{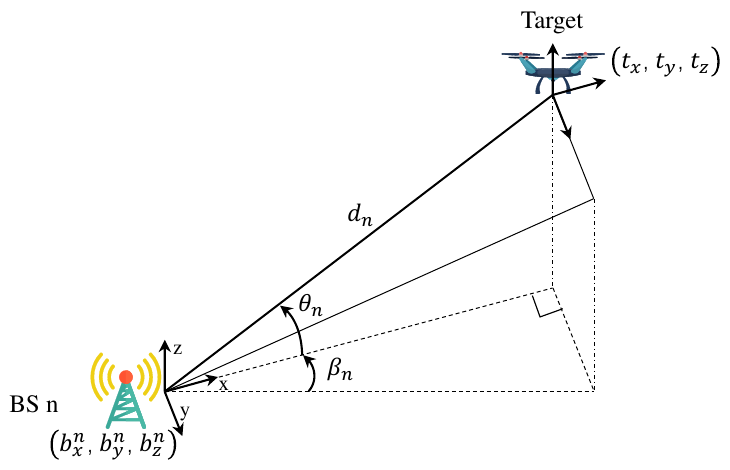} 
\par\end{centering}
\caption{Illustration of the geometric relationship between the sensing target
and a BS.}
\label{fig:sensing_model} 
\end{figure}

Before diving into the signal model for target localization, we first
introduce the geometric settings. Let Cartesian coordinates $\boldsymbol{t}=\left[t_{x},t_{y},t_{z}\right]^{T}$
and $\boldsymbol{b}^{n}=\left[b_{x}^{n},b_{y}^{n},b_{z}^{n}\right]^{T}$
represent the position of the sensing target and the position of the
$n$-th BS, respectively. We assume all BSs are with same height and
placed onto the same plane.As described in Fig. \ref{fig:sensing_model},
the elevation and azimuth angle of the target corresponding to the
$n$-th BS are denoted by $\theta_{n}\in\left[-\pi/2,\pi/2\right]$
and $\beta_{n}\in[-\pi,\pi]$, respectively.Thus, the elevation angle
$\theta_{n}$ can be expressed as
\begin{align}
\theta_{n}=\arcsin\frac{t_{z}-b_{z}^{n}}{d_{n}},\label{eq:theta}
\end{align}
where $d_{n}=\sqrt{\left(t_{x}-b_{x}^{n}\right)^{2}+\left(t_{y}-b_{y}^{n}\right)^{2}+\left(t_{z}-b_{z}^{n}\right)^{2}}$
denotes the Euclidean distance between the $n$-th BS and the sensing
target. The azimuth angle $\beta_{n}$ can be expressed as
\begin{align}
\beta_{n}=\arctan\frac{t_{y}-b_{y}^{n}}{t_{x}-b_{x}^{n}}+\pi\cdot1\left(t_{x}<b_{x}^{n}\right).\label{eq:beta}
\end{align}
If $t_{x}<b_{x}^{n}$, then $1(\cdot)$ equals to $1$, otherwise
$0$.

\begin{figure}[htbp]
\begin{centering}
\includegraphics[width=0.35\textwidth]{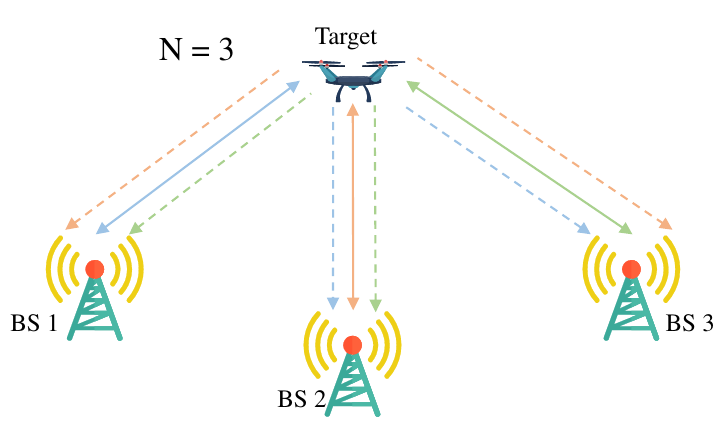} 
\par\end{centering}
\caption{Cooperative localization between multiple base stations.}
\label{fig::LoS} 
\end{figure}

As shown in Fig. \ref{fig::LoS}, we consider that all BSs have line-of-sight
(LoS) links to the sensing target. As a result, the received echo
at the $n$-th BS contains the reflected signal of all BSs by the
target, which is given by
\begin{align}
\boldsymbol{y}_{n} & =\sum_{m=1}^{N}\boldsymbol{H}_{m,n}\boldsymbol{x}_{m}+\boldsymbol{n}_{n}\nonumber \\
 & =\sum_{m=1}^{N}\alpha_{m,n}{\boldsymbol{a}}_{r}\left(\theta_{n},\beta_{n}\right){\boldsymbol{a}}_{t}^{H}\left(\theta_{m},\beta_{m}\right)\mathbf{P}_{m}\mathbf{s}+\boldsymbol{n}_{n},\label{eq::received_n}
\end{align}
where $\boldsymbol{H}_{m,n}$ is the sensing channel response matrix
with respect to the $m$-th transmit BS and the $n$-th receiving
BS, $\alpha_{m,n}$ is the corresponding reflection coefficient, and
$\boldsymbol{n}_{n}\sim\mathcal{CN}(0,\sigma_{n}^{2}\mathbf{I}_{L})$
is the complex Gaussian noise. For simplicity, we consider the BSs
share a common noise level, i.e., $\sigma_{n}^{2}=\sigma^{2}$. Assume
the UPA at BS is placed on the x-z plane with the size $L_{x}\times L_{z}.$
The transmit steering vector $\boldsymbol{a}_{t}\left(\theta_{n},\beta_{n}\right)\in\mathbb{C}^{L\times1}$
and the receive steering vector $\boldsymbol{a}_{r}\left(\theta_{n},\beta_{n}\right)\in\mathbb{C}^{L\times1}$
of a UPA can be expressed as
\begin{align}
\boldsymbol{a}\left(\theta_{n},\beta_{n}\right) & =\frac{1}{\sqrt{L}}\left[1,\cdots,e^{j\frac{2\pi}{\lambda}d\left(i_{x}\cos\beta_{n}\cos\theta_{n}+i_{z}\sin\theta_{n}\right)},\right.\nonumber \\
 & \left.\cdots,e^{j\frac{2\pi}{\lambda}d\left(\left(L_{x}-1\right)\cos\beta_{n}\cos\theta_{n}+\left(L_{z}-1\right)\sin\theta_{n}\right)}\right]^{T},\label{eq::streeing_vector}
\end{align}
where $d$ is the antenna spacing, $\lambda$ is the signal wavelength,
$i_{x}\in\{1,\cdots,L_{x}-1\}$ and $i_{z}\in\{1,\cdots,L_{z}-1\}$
are the antenna indices along the x coordinate and z coordinate. By
reformulating eq.\eqref{eq::received_n} into a matrix form, we derive
\begin{align}
\boldsymbol{y}_{n}={\boldsymbol{a}}_{r}\left(\theta_{n},\beta_{n}\right)\mathbf{A}_{t}(\boldsymbol{\theta,\beta})\mathbf{B}(\boldsymbol{\alpha}_{n})\mathbf{P}\mathbf{s}+\boldsymbol{n}_{n},\label{eq::received_n_matrix}
\end{align}
where
\begin{align}
 & \mathbf{A}_{t}(\boldsymbol{\theta,\beta})=\left[{\boldsymbol{a}}_{t}^{H}\left(\theta_{1},\beta_{1}\right),\cdots,{\boldsymbol{a}}_{t}^{H}\left(\theta_{N},\beta_{N}\right)\right]\in\mathbb{C}^{1\times NL},\\
 & \mathbf{B}(\boldsymbol{\alpha}_{n})=\mathrm{Diag}(\boldsymbol{\alpha}_{n})\otimes\mathbf{I}_{L}\in\mathbb{C}^{NL\times NL},\\
 & \mathbf{P}=\left[\mathbf{P}_{1}^{H},\cdots,\mathbf{P}_{N}^{H}\right]^{H}\in\mathbb{C}^{NL\times(K+1)},\\
 & \boldsymbol{\alpha}_{n}=[{\alpha}_{1,n},...,{\alpha}_{K,n}]^{T},\\
 & \boldsymbol{\theta}=[{\theta}_{1},...,{\theta}_{N}]^{T},\\
 & \boldsymbol{\beta}=[{\beta}_{1},...,{\beta}_{N}]^{T}.
\end{align}

By stacking the received echo signals at each BS as $\boldsymbol{y}=[\boldsymbol{y}_{1}^{H},\cdots,\boldsymbol{y}_{N}^{H}]^{H}$,
we can equivalently get the following expression:
\begin{align}
\boldsymbol{y} & =\mathbf{A}(\boldsymbol{\theta,\beta})\odot\mathbf{B}(\boldsymbol{\alpha})\mathbf{P}\mathbf{s}+\boldsymbol{n},\label{eq:y_matrix}
\end{align}
where $\mathbf{A}(\boldsymbol{\theta,\beta})\in\mathbb{C}^{NL\times NL}$
and $\mathbf{B}(\boldsymbol{\alpha})\in\mathbb{C}^{NL\times NL}$
are given by
\begin{align}
 & \mathbf{A}(\boldsymbol{\theta,\beta})=\left[{\boldsymbol{a}}_{r}^{H}\left(\theta_{1},\beta_{1}\right),\cdots,{\boldsymbol{a}}_{r}^{H}\left(\theta_{N},\beta_{N}\right)\right]^{H}\mathbf{A}_{t}(\boldsymbol{\theta,\beta}),\\
 & \mathbf{B}(\boldsymbol{\alpha})=[\boldsymbol{\alpha}_{1},\cdots,\boldsymbol{\alpha}_{N}]^{T}\otimes\mathbf{1}_{L}.
\end{align}

\subsection{Performance Bound for Cooperative Localization}

\label{subsec:sen_bound} In this subsection, we introduce the SPEB
to evaluate the performance of cooperative localization. SPEB is a
lower bound of mean square error for any unbiased estimator $\hat{\boldsymbol{t}}$
of position $\boldsymbol{t}$ \cite{zhao2020beamspace}, and can be
calculated through
\begin{align}
\mathbb{E}\left\{ \left(\hat{\boldsymbol{t}}-\boldsymbol{t}\right)\left(\hat{\boldsymbol{t}}-\boldsymbol{t}\right)^{T}\right\} \geq\textrm{SPEB}\left(\mathbf{P};\boldsymbol{t}\right)=\textrm{tr}\left(\mathbf{J}_{}^{-1}\left(\mathbf{P};\boldsymbol{t}\right)\right),\label{eq:SPEB}
\end{align}
where $\mathbf{J}$ denotes the Fisher information matrix (FIM) for
position parameter $\boldsymbol{t}$ and $\mathbf{P}$ is the joint
beamforming matrix to be optimized. Based on \cite{gao2023SPEB},
$\mathbf{J}$ can be obtained by firstly calculating the equivalent
FIM of angular parameters $\boldsymbol{\omega}=[\boldsymbol{\theta}^{T},\boldsymbol{\beta}^{T}]^{T}$
, denoted by $\mathbf{J}_{e}\left(\mathbf{P};\boldsymbol{\omega}\right)$,
then transforming it to the FIM of position by multiplying both sides
of $\mathbf{J}_{e}\left(\mathbf{P};\boldsymbol{\omega}\right)$ with
the Jacobian matrix. Specifically, $\mathbf{J}_{e}\left(\mathbf{P};\boldsymbol{\omega}\right)$
can be calculated by
\begin{align}
\mathbf{J}_{e}\left(\mathbf{P};\boldsymbol{\omega}\right)=\frac{2}{\sigma^{2}}\textrm{Re}\left\{ \bar{\mathbf{P}}_{s}^{H}\mathbf{D}^{H}\mathbf{\mathbf{\Pi_{\tilde{A}}^{\perp}}D}\bar{\mathbf{P}}_{s}+\sum_{k=1}^{K}\bar{\mathbf{P}}_{c}^{H}\mathbf{D}^{H}\mathbf{\mathbf{\Pi_{\tilde{A}}^{\perp}}D}\bar{\mathbf{P}}_{c}\right\} ,\label{eq:EFIM}
\end{align}
where the components of \eqref{eq:EFIM} are expressed as
\begin{align}
\mathbf{\Pi_{\tilde{A}}^{\perp}} & =\mathbf{I}-\mathbf{\tilde{A}}\left(\mathbf{\tilde{A}}^{H}\mathbf{\tilde{A}}\right)^{-1}\mathbf{\tilde{A}}^{H}\in\mathbb{C}^{NL\times NL},\\
\tilde{\mathbf{A}} & =\left(\mathbf{1}_{N^{2}}^{T}\otimes\mathbf{A}\right)\odot\frac{\partial\mathbf{B}}{\partial\boldsymbol{\alpha}}\in\mathbb{C}^{NL\times N^{3}L},\\
\frac{\partial\mathbf{B}}{\partial\boldsymbol{\alpha}} & =\left[\frac{\partial\mathbf{B}}{\partial\alpha_{11}},\cdots\frac{\partial\mathbf{B}}{\partial\alpha_{1N}},\cdots\frac{\partial\mathbf{B}}{\partial\alpha_{NN}}\right],\\
\bar{\mathbf{P}}_{s} & =\text{\ensuremath{\mathbf{I}}}_{2N}\otimes\boldsymbol{p}^{s}\in\mathbb{C}^{2N^{2}L\times2N},\\
\bar{\mathbf{P}}_{k} & =\text{\ensuremath{\mathbf{I}}}_{2N}\otimes\boldsymbol{p}_{k}^{c}\in\mathbb{C}^{2N^{2}L\times2N}.
\end{align}
$\mathbf{D}=\left[\mathbf{D}_{\theta},\mathbf{D}_{\beta}\right]\in\mathbb{C}^{NL\times2N^{2}L}$
in \eqref{eq:EFIM} is the partial derivative matrix with respect
to $\boldsymbol{\omega}$. $\mathbf{D}_{\theta}$ and $\mathbf{D}_{\beta}$
are respectively given by
\begin{align}
\mathbf{D}_{\theta}=\left[\mathbf{\frac{\partial\mathbf{A}}{\partial\theta_{1}}}\odot\mathbf{B},\cdots,\mathbf{\frac{\mathbf{\partial A}}{\partial\theta_{N}}}\odot\mathbf{B}\right],\\
\mathbf{D}_{\beta}=\left[\mathbf{\frac{\partial\mathbf{A}}{\partial\beta_{1}}}\odot\mathbf{B},\cdots,\mathbf{\frac{\mathbf{\partial A}}{\partial\beta_{N}}}\odot\mathbf{B}\right].
\end{align}
We further derive the FIM of position $\boldsymbol{t}$ by
\begin{align}
\mathbf{J}\left(\mathbf{P};\boldsymbol{t}\right)=\mathbf{Q}^{T}\mathbf{J}_{e}\left(\mathbf{P};\boldsymbol{\omega}\right)\mathbf{Q},
\end{align}
where $\mathbf{Q}$ is given by
\begin{align}
\mathbf{Q}=\frac{\partial\boldsymbol{\omega}}{\partial\boldsymbol{t}}=\left[\frac{\partial\boldsymbol{\theta}}{\partial\boldsymbol{t}}^{T},\frac{\partial\boldsymbol{\beta}}{\partial\boldsymbol{t}}^{T}\right]^{T},
\end{align}
with
\begin{align}
\frac{\partial\theta_{n}}{\partial\boldsymbol{t}}=\left[-\frac{\left(t_{x}-b_{x}^{n}\right)\left(t_{z}-b_{z}\right)}{d_{n}^{2}d_{xy}^{n}},-\frac{\left(t_{y}-b_{y}^{n}\right)\left(t_{z}-b_{z}\right)}{d_{n}^{2}d_{xy}^{n}},\frac{\left(d_{xy}^{n}\right)^{2}}{d_{n}^{2}d_{xy}^{n}}\right],
\end{align}
and
\begin{align}
\frac{\partial\beta_{n}}{\partial\boldsymbol{t}}=\left[\frac{b_{y}^{n}-t_{y}}{\left(d_{xy}^{n}\right)^{2}},\frac{t_{x}-b_{x}^{n}}{\left(d_{xy}^{n}\right)^{2}},0\right].
\end{align}
$d_{xy}^{n}=\sqrt{\left(b_{x}^{n}-t_{x}\right)^{2}+\left(b_{y}^{n}-t_{y}\right)^{2}}$
denotes the Euclidean distance from the $n$-th BS to the sensing
target on the x-y plane. The computational details can be found in
Appendix \ref{subsec:Derivation-of}.

\subsection{Problem Formulation}

\label{subsec:Problem-Formulation}

To achieve the optimal trade-off between the localization performance
and communication performance in the CoISAC system, we maximize the
total sum-rate of $K$ communication users under the constraints for
localization accuracy and total power by optimizing the joint beamforming
vectors:
\begin{subequations}
\begin{align}
 & \max_{\mathbf{P}}\sum_{k=1}^{K}R_{k}\label{eq:Prob_Obj_func}\\
 & s.t.\;\|\boldsymbol{p}_{n}^{s}\|_{2}^{2}+\sum_{k=1}^{K}\|\boldsymbol{p}_{n,k}^{c}\|_{2}^{2}\leq P_{n},\forall n=1,\cdots,N\label{eq:Con_Power}\\
 & \quad\;\textrm{SPEB}\left(\mathbf{P};\boldsymbol{t}\right)\leq\gamma,\label{eq:Con_radar}
\end{align}
\label{eq: Problem} 
\end{subequations}
where $\gamma$ is the predefined threshold for SPEB. In problem \eqref{eq: Problem},
constraint \eqref{eq:Con_Power} ensures the transmit power limitation
of each BS. Constraint \eqref{eq:Con_radar} guarantees the target
localization performance. Problem \eqref{eq: Problem} is a highly
non-convex problem. Traditional optimization methods rely on a series
of convex approximation techniques to iteratively search for a sub-optimal
solution, which can induce significant communication delays and unreliable
performance in practical systems.

It can be seen that the optimization of beamforming vectors depends
on the CSI of each communication user, i.e., $\boldsymbol{h_{k}}$,
$\forall k\in\mathcal{K}$, through the communication data rate $R_{k}$,
and the position of the target, i.e., $\boldsymbol{t}$, through the
SPEB. Although there are many mature channel estimation algorithms
in communication field and positioning algorithms in sensing filed,
the channel $\hat{\boldsymbol{h}}$ and position $\hat{\boldsymbol{t}}$
obtained by the BSs inevitably suffer from estimation errors, i.e.,
$\hat{\boldsymbol{h}}=\boldsymbol{h}+\Delta\boldsymbol{h}$, $\hat{\boldsymbol{t}}=\boldsymbol{t}+\Delta\boldsymbol{t}$.
Therefore, it is critical to consider robust beamforming design with
the presence of the uncertainties of channel and position. There have
been some standardized optimization techniques to address the parameter
uncertainties, such as Robust Optimization (RO) for bounded error
model \cite{ben2009robust} and Stochastic Programming (SP) for Gaussian
error model \cite{shapiro2021lectures}. However, these methods rely
on idealized uncertainty model and often lead to increased algorithm
complexity. Therefore, it is crucial to design more efficient, robust
and scalable beamfoming algorithms in CoISAC systems. 

\section{Cooperative Beamforming in CoISAC Systems based on LHGNN}

\label{sec:algo_trad} In this section, we model the CoISAC system
using heterogeneous graphs, where different types of links are modeled
as different types of nodes in graph data to effectively capture the
heterogeneous topology and complex interactions of CoISAC systems.
Then we introduce the detailed LHGNN architecture as well as the loss
function to address the joint beamforming optimization problem in
a data-driven manner. Let $\mathcal{Z}=\{1,...,Z\}$ denote the indices
of sensing targets. For simplicity, the problem formulation in Section
\ref{subsec:Problem-Formulation} assumes a single sensing target,
i.e., $Z=1$. However, the proposed algorithm can be easily extended
to a multi-target sensing scenario. Therefore, we present the algorithm
for a general multi-target system, with subscripts involving the target
index $z$ omitted when $Z=1$.

\subsection{Heterogeneous Graph Modeling of the CoISAC Systems}

Heterogeneous graph data can typically be represented as $\mathcal{G}=\left(\mathcal{V,E,O,R}\right)$,
where $\mathcal{V}$ denotes the set of nodes in the graph data, $\mathcal{E}$
represents the set of edges, and sets $\mathcal{O}$ and $\mathcal{R}$
represent the types of nodes and edges, respectively. In the modeling
of the CoISAC system as heterogeneous graph data, we consider a more
general scenario, wherein the modeled system includes $N$ base stations,
$K$ communication users, and $Z$ sensing targets.

\subsubsection{Heterogeneous Node}

When generating the node set $\mathcal{V}$, two types of nodes are
constructed, i.e., $\mathcal{O}=\{o_{c},o_{s}\}$, where $o_{c}$
denotes the communication type, and $o_{s}$ denotes the sensing type,
respectively. To fully capture the heterogeneous nature of CoISAC
systems, we propose an innovative approach that consolidates different
link types in CoISAC into corresponding node types in the graph modeling.
Specifically, there are $NK$ communication links, each indexed by
$(n,k)$, representing the communication link between BS $n$ and
communication user $k$. Additionally, there are $NZ$ sensing links,
each indexed by $(n,z)$, representing the sensing link between BS
$n$ and sensing target $z$. The communication links $\{(n,k),\forall n\in\mathcal{N},k\in\mathcal{K}\}$
are modeled as communication type nodes, and the sensing links $\{(n,z),\forall n\in\mathcal{N},z\in\mathcal{Z}\}$
are modeled as sensing type nodes. The total set of node indices $\mathcal{V}=\{1,2,\cdots,NK+NZ\}$
contains $NK$ communication type nodes and $NZ$ sensing type nodes.
The mapping between a link $(n,k)$ (or $(n,z)$) and its corresponding
node index $v\in\mathcal{V}$ is represented by function $\varphi$,
such that $\varphi(n,\cdot)=v,$ with the inverse mapping given by
$\varphi(v)^{-1}=(n,\cdot)$.

\subsubsection{Heterogeneous Edge}

When generating the edge set $\mathcal{E}$ for the heterogeneous
graph data, neighbor nodes are constructed separately for the two
node types, i.e., $o_{c}$ and $o_{s}$, thereby completing the construction
of the edge set. The set of neighboring nodes for node $v\in\mathcal{V}$
is defined as
\begin{align}
\mathcal{N}(v)=\{j\in\mathcal{V}:(v,j)\in\mathcal{E}\}.
\end{align}
Next, we will explain how to construct the neighboring nodes based
on the type of the central node.

\begin{figure}[htbp]
\begin{centering}
\includegraphics[width=0.38\textwidth]{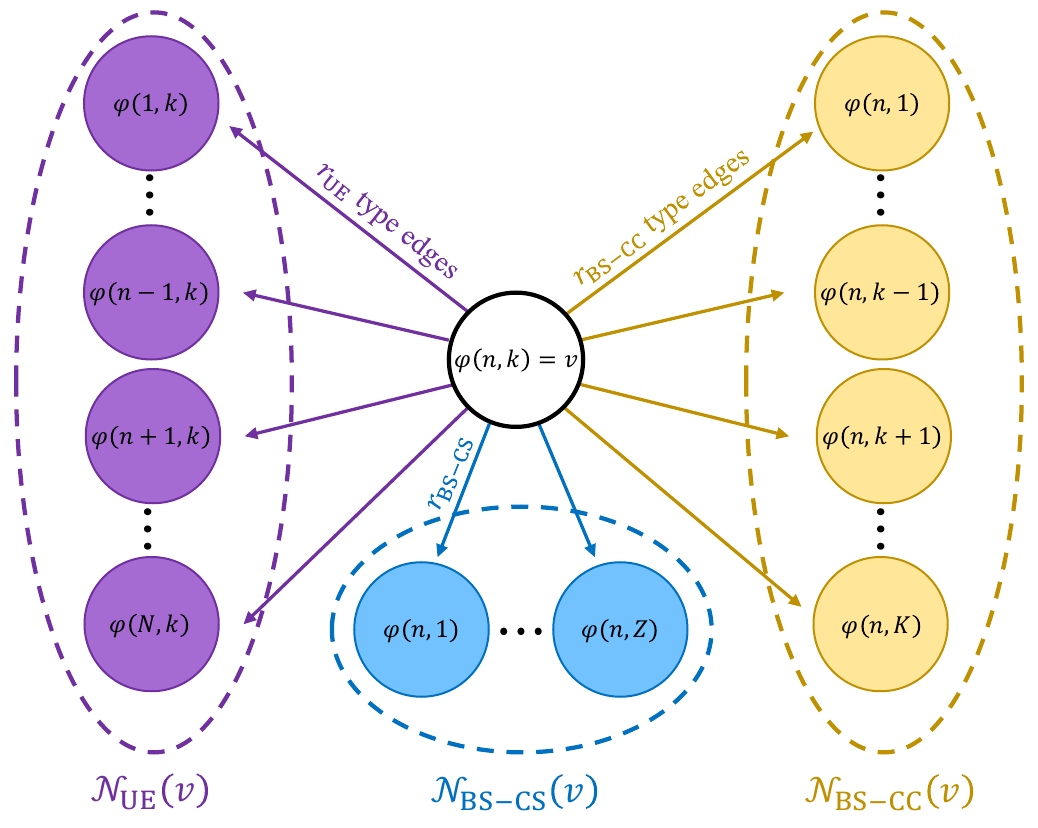} 
\par\end{centering}
\caption{A typical node of type $\boldsymbol{o_{c}}$ and its three types of
neighbors.}
\label{img::nk_neighbor} 
\end{figure}

For a central node $\varphi(n,k)$ of type $o_{c}$, its connections
to neighbor nodes include three different types of edges, namely $\mathcal{R}_{c}=\{r_{{\rm BS-CC}},r_{{\rm UE}},r_{{\rm BS-CS}}\}$.
If a neighbor node of type $o_{c}$ shares the same base station with
the central node, the edge connecting them is of type $r_{{\rm BS-CC}}$.
If a neighbor node of type $o_{c}$ shares the same communication
user with the central node, the edge connecting them is of type $r_{{\rm UE}}$.
If a neighbor node of type $o_{s}$ shares the same base station with
the central node, the edge connecting them is of type $r_{{\rm BS-CS}}$.
Define ${\rm type}(\cdot)$ as the mapping from an edge $e\in\mathcal{E}$
to an edge type $r\in\mathcal{R}$. The construction of these three
types of edges can be summarized by the following three expressions:
\begin{align}
e=\left(\varphi(n,k),\varphi(n,k^{\prime})\right),{\rm type}(e) & =r_{{\rm BS-CC}},\\
e=\left(\varphi(n,k),\varphi(n^{\prime},k)\right),{\rm type}(e) & =r_{{\rm UE}},\\
e=\left(\varphi(n,k),\varphi(n,z)\right),{\rm type}(e) & =r_{{\rm BS-CS}}.
\end{align}
Therefore, three different types of neighbor node sets for the central
node $\varphi(n,k)=v$ can be represented as $\mathcal{N}_{BS-CC}(v)$,
$\mathcal{N}_{UE}(v)$, and $\mathcal{N}_{{\rm BS-CS}}(v)$, as shown
in Fig. \ref{img::nk_neighbor}. These three neighbor sets satisfy
the following properties:
\begin{align}
 & \mathcal{N}_{{\rm BS-CC}}(v)\cup\mathcal{N}_{{\rm UE}}(v)\cup\mathcal{N}_{{\rm BS-CS}}(v)=\mathcal{N}(v),\\
 & \mathcal{N}_{{\rm BS-CC}}(v)\cap\mathcal{N}_{{\rm UE}}(v)=\mathcal{N}_{{\rm BS-CC}}(v)\cap\mathcal{N}_{{\rm BS-CS}}(v)\nonumber \\
 & =\mathcal{N}_{{\rm UE}}(v)\cap\mathcal{N}_{{\rm BS-CS}}(v)=\emptyset.\nonumber 
\end{align}
Each $\mathcal{N}_{{\rm BS-CC}}(v)$ set contains $K-1$ elements,
each $\mathcal{N}_{{\rm UE}}(v)$ set contains $N-1$ elements, and
each $\mathcal{N}_{{\rm BS-CS}}(v)$ set contains $Z$ elements, respectively.
Therefore, each $\mathcal{N}(v)$ set contains $K+N+Z-2$ elements,
when $v$ is of type $o_{c}$ .

\begin{figure}[htbp]
\begin{centering}
\includegraphics[width=0.38\textwidth]{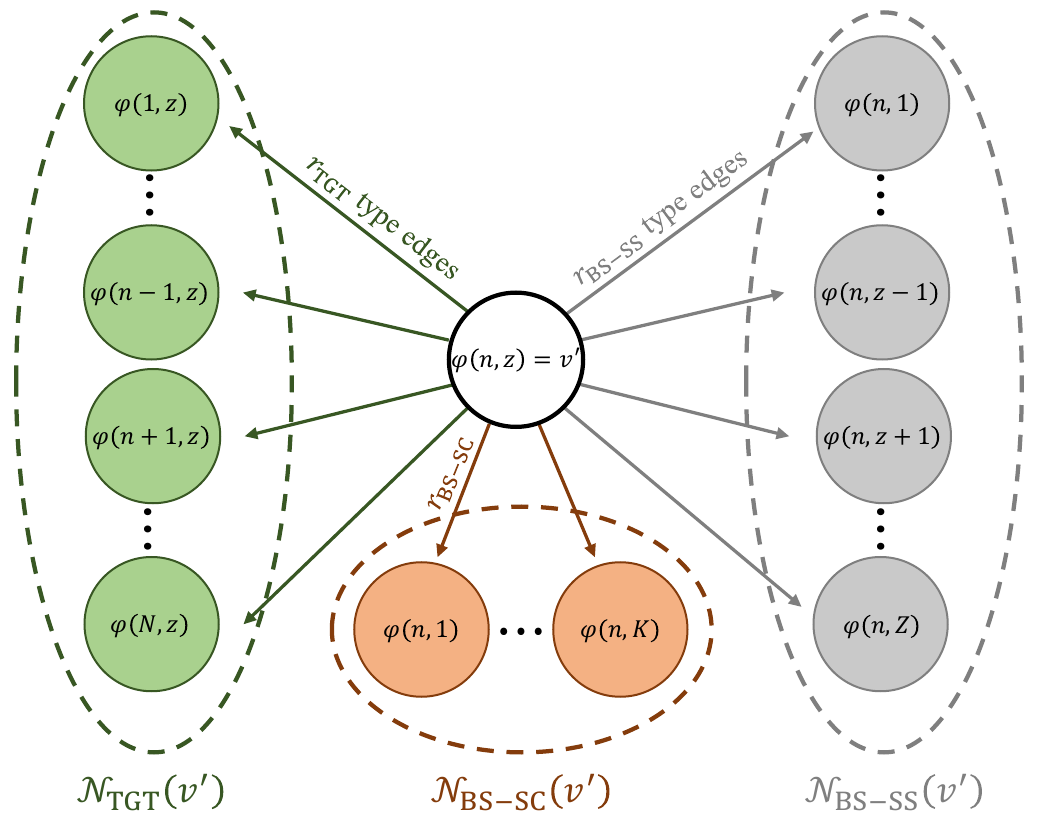} 
\par\end{centering}
\caption{A typical node of type $\boldsymbol{o_{s}}$ and its three types of
neighbors.}
\label{img::nz_neighbor}
\end{figure}

For a central node $\varphi(n,z)$ of type $o_{s}$, the connections
to its neighbor nodes also include three different types of edges,
namely $\mathcal{R}_{s}=\{r_{{\rm BS-SS}},r_{{\rm TGT}},r_{{\rm BS-SC}}\}$.
If a neighbor node of type $o_{s}$ shares the same base station with
the central node, the edge connecting them is of type $r_{{\rm BS-SS}}$.
If a neighbor node of type $o_{s}$ shares the same sensing target
with the center node, the edge connecting them is of type $r_{{\rm TGT}}$.
If a neighbor node of type $o_{c}$ shares the same BS with the central
node, the edge connecting them is of type $r_{{\rm BS-SC}}$. The
construction of these three types of edges can be summarized as
\begin{align}
e=\left(\varphi(n,z),\varphi(n,z^{\prime})\right),{\rm type}(e) & =r_{{\rm BS-SS}},\\
e=\left(\varphi(n,z),\varphi(n^{\prime},z)\right),{\rm type}(e) & =r_{{\rm TGT}},\\
e=\left(\varphi(n,z),\varphi(n,k)\right),{\rm type}(e) & =r_{{\rm BS-SC}}.
\end{align}
Therefore, the three types of neighbor node sets for the central node
$\varphi(n,z)=v^{\prime}$ can be represented as $\mathcal{N}_{{\rm BS-SS}}(v^{\prime})$,
$\mathcal{N}_{{\rm TGT}}(v^{\prime})$ and $\mathcal{N}_{{\rm BS-SC}}(v^{\prime})$,
as shown in Fig. \ref{img::nz_neighbor}. These three neighbor sets
satisfy the following properties:
\begin{align}
 & \mathcal{N}_{{\rm BS-SS}}(v^{\prime})\cup\mathcal{N}_{{\rm TGT-SS}}(v^{\prime})\cup\mathcal{N}_{{\rm BS-SC}}(v)=\mathcal{N}(v^{\prime}),\\
 & \mathcal{N}_{{\rm BS-SS}}(v^{\prime})\cap\mathcal{N}_{{\rm TGT}}(v^{\prime})=\mathcal{N}_{{\rm BS-SS}}(v^{\prime})\cap\mathcal{N}_{{\rm BS-SC}}(v^{\prime})\nonumber \\
 & =\mathcal{N}_{{\rm TGT}}(v^{\prime})\cap\mathcal{N}_{{\rm BS-SC}}(v^{\prime})=\emptyset,
\end{align}
Each $\mathcal{N}_{{\rm BS-SS}}(v^{\prime})$ set contains $Z-1$
elements, each $\mathcal{N}_{{\rm TGT}}(v^{\prime})$ set contains
$N-1$ elements, and each $\mathcal{N}_{{\rm BS-SC}}(v^{\prime})$
set contains $K$ elements, respectively. Therefore, each $\mathcal{N}(v^{\prime})$
set contains $K+N+Z-2$ elements, when node $v'$ is of type $o_{s}$.

Note that the construction of different types of edges in the graph
depends on the joint beamforming design problem \eqref{eq: Problem}
in the CoISAC systems. For example, due to the per-BS power constraints,
the performance of the communication and sensing links connected to
the same BS is mutually constrained. Meanwhile, in the expressions
of communication rate \eqref{eq:rate} and sensing SPEB \eqref{eq:SPEB},
the communication beams and sensing beams transmitted from the same
BS exhibit complex interactions, such as interference or mutual enhancement,
leading to performance trade-offs between communication and sensing
links associated with the same BS. Consequently, it is necessary to
construct the edge types, such as $r_{{\rm BS-CC}}$, $r_{{\rm BS-CS}}$,
$r_{{\rm BS-SS}}$ and $r_{{\rm BS-SC}}$, to model these interactions.
Furthermore, in the CoISAC systems, multi-BS cooperation enables collaborative
communication with the same user and joint positioning of the same
target, thereby enhancing the communication and sensing performance.
This requires the construction of $r_{{\rm UE}}$ and $r_{{\rm TGT}}$
types of edges to reflect such cooperative relationships.Heterogeneous
Graph

For a CoISAC system with $N$ BSs, $K$ users, and $Z$ sensing targets,
a heterogeneous graph structure can be constructed using the aforementioned
methods. This heterogeneous graph contains $NK$ $o_{c}$-type nodes
and $NZ$ $o_{s}$-type nodes. Each $o_{c}$-type node is connected
to its $K+N+Z-2$ neighbor nodes through three types of edges, i.e.,
$\{r_{{\rm BS-CC}},r_{{\rm UE}},r_{{\rm BS-CS}}\}$, while each $o_{s}$-type
node is connected to its $K+N+Z-2$ neighbor nodes through three types
of edges, i.e., $\{r_{{\rm BS-SS}},r_{{\rm TGT}},r_{{\rm BS-SC}}\}$.
Note that when there is only one sensing target, edges of type $r_{{\rm BS-SS}}$
will not exist. In the next subsection, we will explain how to perform
learning over the constructed heterogeneous graph.

\subsection{Framework of Attention-Enhanced LHGNN}

\begin{figure*}[t]
\begin{centering}
\includegraphics[width=0.9\textwidth]{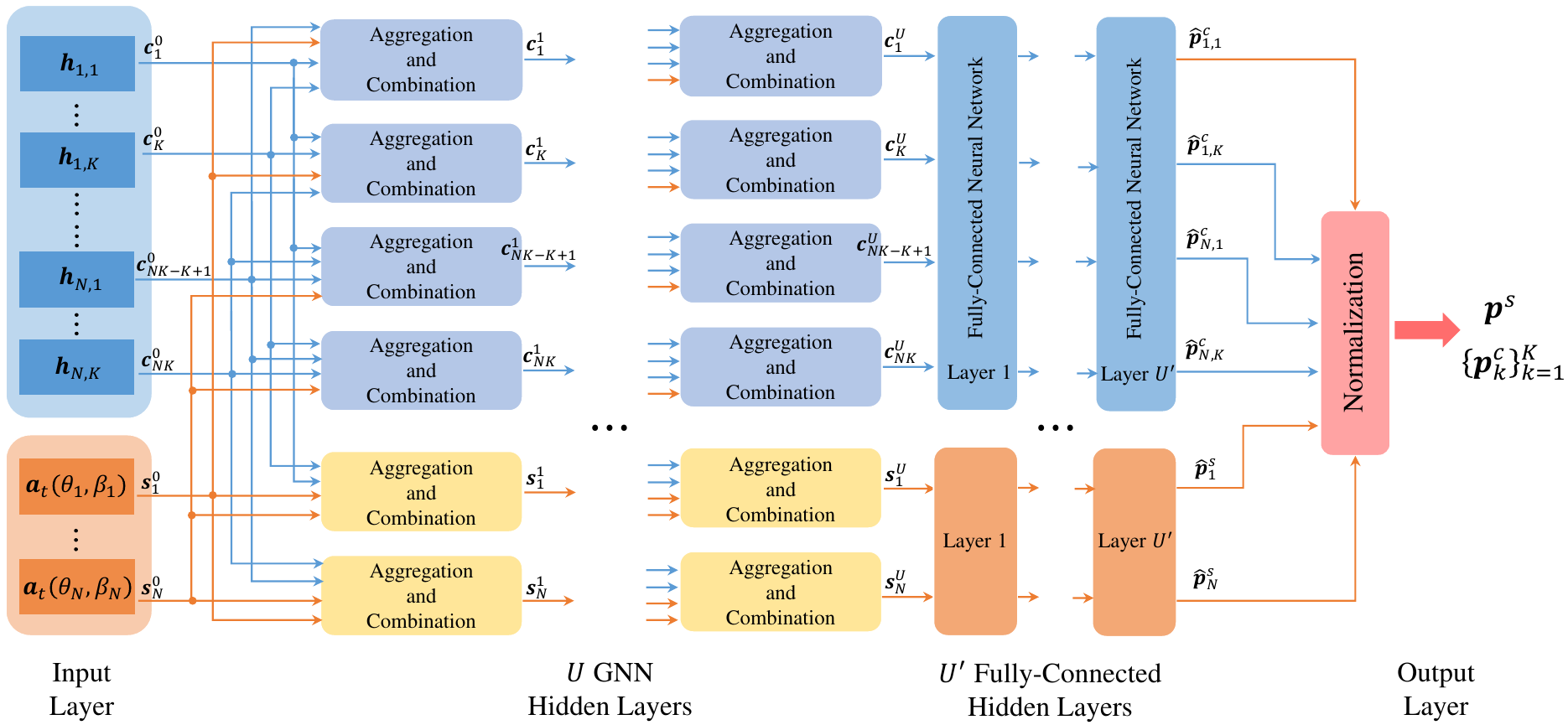} 
\par\end{centering}
\caption{The overall structure of the proposed heterogeneous graph neural network.}
\label{img::GNN_arcitecture} 
\end{figure*}

The LHGNN performs learning on the proposed heterogeneous graph by
leveraging information from neighbor nodes to enrich individual node
features and propagating these features throughout the graph following
its structured connections. Integrating an attention mechanism into
the LHGNN introduces learnable weights during information aggregation,
thereby enhancing the robustness of the network to the input data
errors. In this subsection, we will provide a detailed introduction
to the proposed LHGNN architecture enhanced by attention mechanism,
comprising four components as shown in Fig. \ref{img::GNN_arcitecture}:
an input layer, $U$ GNN hidden layers, $U^{\prime}$ fully connected
hidden layers, and an output layer with normalization operation. The
input layer is utilized for embedding the input data, followed by
the GNN hidden layers, which aggregate and update node features across
the heterogeneous graph. During this process, an attention mechanism
is employed to effectively and robustly extract the features based
on the topological structure of CoISAC systems. The fully connected
hidden layers learn the mapping between the graphical features and
the joint beamforming vector for each link. Finally, the output layer
generates the joint beamforming vectors corresponding to each link
node.

\subsubsection{Input Layer}

In the input layer, initial node features are constructed for nodes
of type $o_{c}$ representing communication links and nodes of type
$o_{s}$ representing sensing links. These features serve as the input
data for the GNN. For node $i=\varphi(n,k),\forall n\in\mathcal{N},k\in\mathcal{K}$
of type $o_{c}$, the CSI $\boldsymbol{h}_{n,k}$ of the corresponding
link $(n,k)$ is used as the initial node feature. Therefore, the
initial feature $c_{i}^{0}$ for node $i$ corresponding to communication
link $(n,k)$ is: 
\begin{align}
\boldsymbol{c}_{i}^{0}=\left[{\rm Re}(\boldsymbol{h}_{k,n}),{\rm Im}(\boldsymbol{h}_{k,n})\right]^{T}\in\mathbb{R}^{2L\times1}.
\end{align}
For node $i=\varphi(n,z),\forall n\in\mathcal{N},z\in\mathcal{Z}$
of type $o_{s}$, the transmit steering vector $\boldsymbol{a}_{t}(\theta_{n,z},\beta_{n,z})$
of the corresponding link $(n,z)$ is used as the initial node feature,
as shown in equation \eqref{eq::streeing_vector}, where the elevation
angle $\theta_{n,z}$ and azimuth angle $\beta_{n,z}$ are related
to the position $\boldsymbol{t_{z}}$ of the target $z$ through the
geometric relationships. Therefore, the initial feature $s_{i}^{0}$
for node $i$ corresponding to sensing link $(n,z)$ is: 
\begin{align}
\boldsymbol{s}_{i}^{0}=\left[{\rm Re}\left(\boldsymbol{a}_{t}(\theta_{n,z},\beta_{n,z})\right),{\rm Im}\left(\boldsymbol{a}_{t}(\theta_{n,z},\beta_{n,z})\right)\right]^{T}\in\mathbb{R}^{2L\times1}.
\end{align}
Note that the initial node features depend on the CSI of communication
links and the position of sensing target, which can suffer from estimation
errors. Therefore, during training, the input data for the network
are the noisy CSI estimates $\hat{\boldsymbol{h}}_{k,n}$ and the
noisy angles $\hat{\theta}_{n,z}$ and $\hat{\beta}_{n,z}$, such
that the network can automatically learn how to mitigate the impact
of input parameter errors during graphical feature extraction and
beamforming learning, thus improving the robustness of the network.

\subsubsection{GNN Hidden Layer}

In the $U$ GNN hidden layers, the network aggregates and updates
feature information between neighboring nodes based on the constructed
heterogeneous graph data, effectively utilizing topological information.
Due to heterogeneity, different neural networks are used for feature
mapping for different types of nodes and edges during the feature
aggregation. To boost the robustness of the proposed method against
estimation errors, an attention mechanism is introduced. The feature
update process for the $i$-th node of type $o_{c}$ in the $u$-th
GNN hidden layer is represented as follows: 
\[
\boldsymbol{c}_{i}^{u+1}=\frac{1}{|\mathcal{N}(i)|}\sum_{j\in\mathcal{N}(i)}{\rm Norm}\left(\mathcal{F}_{\mathcal{J}}^{u+1}\right)\in\mathbb{R}^{L^{u+1}\times1},
\]
\begin{align}
\boldsymbol{c}_{i}^{u+1}=\frac{1}{|\mathcal{R}_{c}|}\sum_{r_{\bullet}\in\mathcal{R}_{c}}{\rm Norm}\left(\mathcal{F}_{r_{\bullet}}^{u+1}\right)\in\mathbb{R}^{L^{u+1}\times1},
\end{align}
where $\mathcal{R}_{c}=\{r_{{\rm BS-CC}},r_{{\rm UE}},r_{{\rm BS-CS}}\}$
represents the set of different types of edges for nodes of type $o_{c}$,
${\rm Norm}(\cdot)$ represents the normalization function layer,
$L^{u+1}$ denotes the dimension of the feature $\boldsymbol{c}_{i}^{u+1}$
for node $i$ after the update, and $\mathcal{F}_{r_{\bullet}}^{u+1}$
represents the aggregated features of type $r_{\bullet}$ connecting
edges, calculated as follows: 
\begin{align}
\mathcal{F}_{r_{\bullet}}^{u+1} & =f_{\bullet}^{1,u}(\boldsymbol{c}_{i}^{u})+\frac{1}{|\mathcal{N}_{\bullet}(i)|}\sum_{j\in\mathcal{N}_{\bullet}(i)}\lambda_{\bullet}^{u+1}(i,j)\times f_{\bullet}^{2,u}(\boldsymbol{c}_{j}^{u}).\label{eq::update_c}
\end{align}
When $\bullet={\rm BS-CS}$, the features of neighbor nodes to be
aggregated in eq. \eqref{eq::update_c} should be changed from $\boldsymbol{c}_{j}^{u}$
to $\boldsymbol{s}_{j}^{u}$. The function $f(\cdot)$ represents
a fully connected layer. Different fully connected networks (FCN)
are used for different hidden GNN layers and for different edge types.
The attention coefficient $\lambda$ is calculated as follows: 
\begin{align}
\lambda_{\bullet}^{u+1}=\frac{\left\langle f_{\bullet}^{3}(\boldsymbol{c}_{i}^{u}),f_{\bullet}^{4}(\boldsymbol{c}_{j}^{u})\right\rangle }{\sum_{j^{\prime}\in\mathcal{N}_{\bullet}(i)}\left\langle f_{\bullet}^{3}(\boldsymbol{c}_{i}^{u}),f_{\bullet}^{4}(\boldsymbol{c}_{j^{\prime}}^{u})\right\rangle },\label{eq::attention}
\end{align}
where $\langle\boldsymbol{x},\boldsymbol{y}\rangle={\rm exp}\left(\frac{\boldsymbol{x}^{T}\boldsymbol{y}}{l}\right)$
represents the exponential scaled dot product \cite{vaswani2017attention},
and $l$ is the dimension of the vector. In expression \eqref{eq::attention},
$l=L^{u+1}$. Similarly, when $\bullet={\rm BS-CS}$, $\boldsymbol{c}_{j}^{u}$
in this expression should be replaced with $\boldsymbol{s}_{j}^{u}$.
The introduction of the attention mechanism allows each node to focus
on a subset of its neighbors, rather than considering all neighbors
equally. Additionally, with learnable attention coefficients, the
aggregation process can assign greater weight to neighbors that are
more significant in the beamforming design, thus enabling the model
to have certain robustness to noise in the initial node features.
Similarly, the feature update process for the $i$-th node of type
$o_{s}$ in the $u$-th GNN hidden layer can be represented as 
\begin{align}
\boldsymbol{s}_{i}^{u+1}=\frac{1}{|\mathcal{R}_{s}|}\sum_{r_{\bullet}\in\mathcal{R}_{s}}{\rm Norm}\left(\mathcal{P}_{r_{\bullet}}^{u+1}\right)\in\mathbb{R}^{L^{u+1}\times1},
\end{align}
where $\mathcal{R}_{s}=\{r_{{\rm TGT}},r_{{\rm BS-SC}},r_{{\rm BS-SS}}\}$
represents the set of different types of edges for nodes of type $o_{s}$,
$\bullet\in\{{\rm TGT},{\rm BS-SC},{\rm BS-SS}\}$, and $\mathcal{P}_{r_{\bullet}}^{u+1}$
represents the aggregated features of type $r_{\bullet}$ connecting
edges: 
\begin{align}
\mathcal{P}_{r_{\bullet}}^{u+1} & =f_{\bullet}^{1}(\boldsymbol{s}_{i}^{u})+\frac{1}{|\mathcal{N}_{\bullet}(i)|}\sum_{j\in\mathcal{N}_{\bullet}(i)}\lambda_{\bullet}^{u+1}(i,j)\times f_{\bullet}^{2}(\boldsymbol{s}_{j}^{u}).\label{eq::update_s}
\end{align}
 When $\bullet={\rm BS-SC}$, features $\boldsymbol{s}_{j}^{u}$ in
eq.\eqref{eq::update_s} should be replaced by $\boldsymbol{c}_{j}^{u}$.
After feature aggregation and updating on $U$ LHGNN hidden layers,
feature vectors $\boldsymbol{c}_{i}^{U}$ and $\boldsymbol{s}_{i}^{U}$
containing the topological structure information in the heterogeneous
graph are obtained at the output of the $U$-th hidden layer.

\subsubsection{Fully-Connected Hidden Layers}

$U^{\prime}$ fully connected hidden layers are utilized to learn
the relationship between the feature vectors $\boldsymbol{c}_{i}^{U}$
and $\boldsymbol{s}_{i}^{U}$ and the beamforming vectors of the corresponding
links. At the output of the $U^{\prime}$-th fully connected hidden
layer, the real and imaginary parts of the communication and sensing
beamforming vectors corresponding to each node's link are obtained.
These are then converted to complex numbers to obtain the joint beamforming
matrix $\mathbf{P}$ required to be solved in the optimization problem
\eqref{eq:Prob_Obj_func}. Specifically: 
\begin{itemize}
\item For node $i=\varphi(n,k)$ of type $o_{c}$, the corresponding communication
beamforming vector $\hat{\boldsymbol{p}}_{n,k}^{c}$ from the $n$-th
base station to the $k$-th user can be obtained through $\left[{\rm Re}(\hat{\boldsymbol{p}}_{n,k}^{c}),{\rm Im}(\hat{\boldsymbol{p}}_{n,k}^{c})\right]^{T}=FL_{c}(\boldsymbol{c}_{i}^{U};W_{c}^{(1)},b_{c}^{(1)},...,W_{c}^{(U')},b_{c}^{(U')})$,
where $FL_{c}(\cdot)$ denotes the composite function of the fully
connected neural network for feature vector $\boldsymbol{c}_{i}^{U}$,
and $W_{c}$ and $b_{c}$ denote the weight matrix and bias of each
layer in the network, respectively.
\item For node $i=\varphi(n,z)$ of type $o_{s}$, the corresponding sensing
beamforming vector $\hat{\boldsymbol{p}}_{n,z}^{s}$ from the $n$-th
base station to the $z$-th sensing target can be obtained through
$\left[{\rm Re}(\hat{\boldsymbol{p}}_{n,z}^{s}),{\rm Im}(\hat{\boldsymbol{p}}_{n,z}^{s})\right]^{T}=FL_{s}(\boldsymbol{s}_{i}^{U};W_{s}^{(1)},b_{s}^{(1)},...,W_{s}^{(U')},b_{s}^{(U')})$,
where $FL_{s}(\cdot)$ denotes the composite function of the fully
connected neural network for feature vector $\boldsymbol{s}_{i}^{U}$,
and $W_{s}$ and $b_{s}$ denote the weight matrix and bias of each
layer in the network, respectively.
\end{itemize}

\subsubsection{Output Layer}

At the output layer, a normalization operation is performed on the
beamforming vectors $\hat{\boldsymbol{p}}_{n,k}^{c}$ and $\hat{\boldsymbol{p}}_{n,z}^{s}$
learned by the fully connected hidden layers. This ensures that the
final output beamforming vectors $\boldsymbol{p}_{n,k}^{c}$ and $\boldsymbol{p}_{n,z}^{s}$
satisfy the power constraint in equation \eqref{eq:Con_Power}. The
specific normalization process is as follows: 
\begin{subequations}
\begin{align}
\boldsymbol{p}_{n,k}^{c}=\sqrt{P_{n}}\frac{\hat{\boldsymbol{p}}_{n,k}^{c}}{{\rm max}(\sqrt{P_{n}},\sqrt{\hat{P}_{n}})},\\
\boldsymbol{p}_{n,z}^{s}=\sqrt{P_{n}}\frac{\hat{\boldsymbol{p}}_{n,z}^{s}}{{\rm max}(\sqrt{P_{n}},\sqrt{\hat{P}_{n}})},\\
\hat{P}_{n}=\sum_{z=1}^{Z}\|\hat{\boldsymbol{p}}_{n,z}^{s}\|_{2}^{2}+\sum_{k=1}^{K}\|\hat{\boldsymbol{p}}_{n,k}^{c}\|_{2}^{2}.
\end{align}
\end{subequations}

\subsection{Loss Design of the LHGNN}

In this subsection, we introduce the loss design of the LHGNN. An
unsupervised learning approach is adopted to train the LHGNN, which
enables the learning of the optimal beamforming vectors without labeled
data. The literature \cite{gill2019practical} demonstrates that by
introducing penalty terms, a constrained optimization problem can
be equivalently transformed into an unconstrained one. In the optimization
problem \eqref{eq:Prob_Obj_func}, power constraint \eqref{eq:Con_Power}
has already been considered in the normalization operation at the
output layer. Therefore, when designing the penalty term of the loss
function, only constraint \eqref{eq:Con_radar} needs to be considered.
Let the training dataset be $\{\mathbf{H}^{i},\mathbf{A}_{t}^{i}(\boldsymbol{\theta,\beta})\}_{i=1}^{I_{T}}$.
The expression for the loss function is as follows: 
\begin{subequations}
\begin{align}
\mathcal{L}= & -\frac{1}{I_{T}}\sum_{i=1}^{I_{T}}\sum_{k=1}^{K}R_{k}^{i}\label{eq::GNN_LF_1}\\
 & +\rho_{1}\frac{1}{I_{T}}\sum_{i=1}^{I_{T}}\left[{\rm max}\left(0,\textrm{SPEB}^{i}-\gamma\right)\right]\label{eq::GNN_LF_2}\\
 & +\rho_{2}\frac{1}{I_{T}}\sum_{i=1}^{I_{T}}\sum_{k=1}^{K}\left[{\rm max}\left(0,r-R_{k}^{i}\right)\right].\label{eq::GNN_LF_3}
\end{align}
\label{eq::GNN_LF} 
\end{subequations}

The entire loss function can be divided into three parts. The first
part is the negative of the objective function. By minimizing this
part, we achieve the maximization of the achievable rate. The second
part considers the constraint \eqref{eq:Con_Power}, ensuring that
the designed joint beamforming matrix can maximize the achievable
rate while also guaranteeing the performance of sensing. The third
part is a penalty term for local optimal solutions. Due to the highly
non-convexity of the problem, the training of LHGNN can converge to
bad local minima. To address this, a threshold is imposed on the per-user
data rate during the model training to avoid some extreme local optima
where the data rate is significantly low.

\section{Simulation Results }\label{sec:Simulation-Results}

In this section, simulation results are provided to demonstrate the
effectiveness of the proposed cooperative beamforming algorithm based
on LHGNN. We compare the sensing performance, communication performance,
and robustness of the proposed algorithm with several benchmarks. 

\subsection{Parameter Settings}

The simulation uses the "O1" outdoor scenario from the DeepMIMO
dataset \cite{alkhateeb2019deepmimo} to generate the required data,
with the center frequency set to 28 GHz. The number of BSs is set
to $N=3$, using BSs with indices 1, 2, and 3 from the dataset. The
coordinates of these three BSs are $\boldsymbol{b}^{1}=\left[236,390,6\right]^{T}$,
$\boldsymbol{b}^{2}=\left[288,390,6\right]^{T}$, and $\boldsymbol{b}^{3}=\left[236,490,6\right]^{T},$
respectively. Each BS is equipped with $L=L_{x}\times L_{y}=16$ antennas,
where $L_{x}=L_{y}=4$. The system includes $K=5$ single-antenna
users The coordinates of the target to be sensed are set to $\boldsymbol{t}=\left[244,456,22\right]$.

The unsupervised learning of the proposed LHGNN is completed using
$I_{T}=4000$ training data points. The LHGNN consists of $U=9$ GNN
hidden layers and $U^{\prime}=2$ fully connected hidden layers. In
each GNN hidden layer, feature mapping, and attention coefficient
learning are accomplished through fully connected networks. The output
dimensions of features for different GNN hidden layers are as follows:
\begin{align*}
\left[2\times L,128,128,128,256,512,256,128,2\times L\right].
\end{align*}

In the loss function, the hyperparameters are set as $\rho_{1}=0.8\times{\rm epoch}$
and $\rho_{2}=0.5\times{\rm epoch}$ respectively, where epoch represents
the current training round. The total number of training rounds is
set to $800$, and the threshold in the penalty term for local optimal
solutions is set to $r=0.05$. The model parameters of the network
are updated and optimized by the Adam optimizer, with the initial
learning rate set to $0.0002$ and the weight decay coefficient set
to $0.0005$.

\subsection{Benchmark Algorithms}

To evaluate the performance of the proposed LHGNN in designing joint
beamforming matrices, the following three methods are used as comparison
benchmarks:

\textit{(1) SRW }\cite{gao2023SPEB}: In this algorithm, an iterative
algorithm based on SCA is designed to complete the joint beamforming
matrix design. In the performance comparison, a strict constraint
is set for SPEB to ensure a certain level of performance for the sensing
task of the ISAC system.

\textit{(2) Homo GNN:} In this algorithm, the heterogeneity of node
and edge types is not considered. A homogeneous GNN is adopted, where
the modeled graph data contains only one type of node and one type
of edge. The design of the adjacency matrix and the initial node features
are consistent with the proposed method. Similarly, a graph attention
mechanism is also considered to automatically learn the aggregation
priorities.

\textit{(3) Naive CNN:} In this algorithm, convolutional neural networks
(CNN) are adopted to extract features, and fully connected layers
are then used to learn the mapping from features to joint beamforming.
The input of the CNN is consistent with the initial node features
of the proposed method.

When training the proposed algorithm, as well as Homo GNN and Naive
CNN, unsupervised learning methods are adopted. The expression of
the loss function is shown in equation \eqref{eq::GNN_LF}, where
the threshold $\gamma$ for sensing performance is set to be the value
of the SPEB result obtained after optimization by the SRW algorithm,
such that all schemes are compared under the same sensing performance
conditions 

\subsection{Communication and Sensing Performance Evaluation}

\begin{figure}[t]
\begin{centering}
\includegraphics[width=0.38\textwidth]{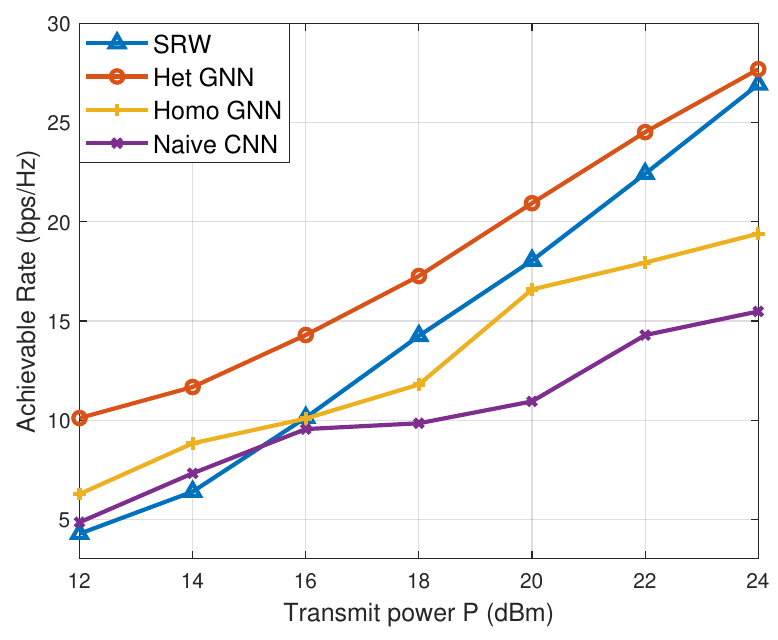} 
\par\end{centering}
\caption{Achievable rate versus transmit power of different algorithms.}
\label{img::Rate_ISAC} 
\end{figure}

To evaluate the performance of different algorithms in terms of the
communication metric of achievable rate, simulations are conducted
under different power constraint settings, as shown in Fig. \ref{img::Rate_ISAC}.
The results show that the achievable rates of all algorithms increase
with the maximum transmission power. The proposed algorithm based
on LHGNN can achieve better communication performance under different
power constraints compared to other algorithms. Compared to the Naive
CNN algorithm, both homogeneous and LHGNN algorithms achieve better
performance. This indicates that by using GNN, features containing
network topology information can be effectively extracted, thereby
achieving better performance of joint beamforming. Compared to the
Homo GNN algorithm, the performance gain of LHGNN comes from modeling
the ISAC system using link heterogeneous graphs. The diversity of
node types and the complicated interactions between communication
links and sensing links during the optimization are better represented
with heterogeneous graphs. Compared to the SRW algorithm, the proposed
algorithm can achieve better communication performance under different
transmission powers and shows greater performance gain when the maximum
transmission power is smaller. 

\begin{figure}[htbp]
\begin{centering}
\includegraphics[width=0.38\textwidth]{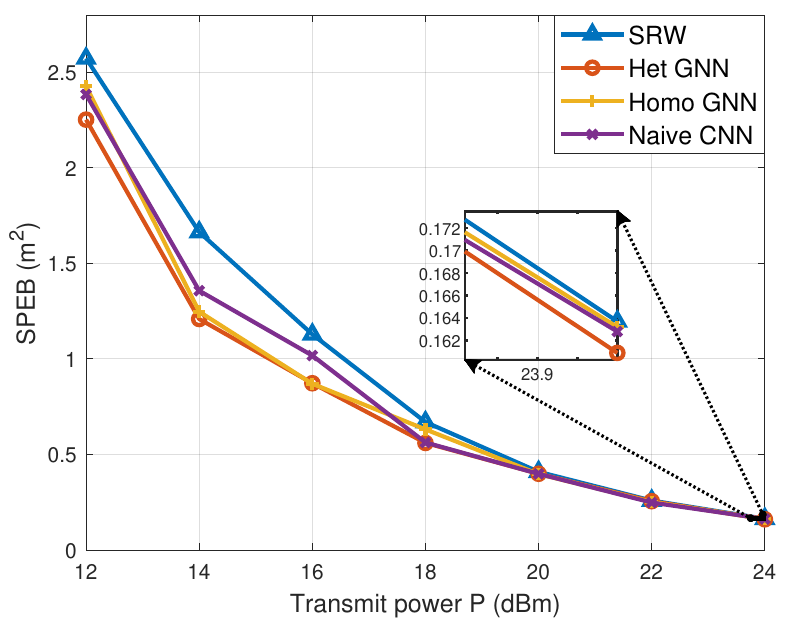} 
\par\end{centering}
\caption{SPEB versus transmit power of different algorithms.}
\label{img::SPEB_ISAC} 
\end{figure}

To verify the effectiveness of the loss function in guaranteeing the
sensing performance, a further comparison of the SPEB under different
maximum transmission power is conducted, as shown in Fig. \ref{img::SPEB_ISAC}.
The threshold $\gamma$ of sensing performance for the learning-based
algorithms is set to the same optimized value as in SRW. The simulation
results show that by using the designed loss function \eqref{eq::GNN_LF},
the joint beamforming matrices learned by different neural networks
can satisfy the constraint \eqref{eq:Con_radar} in the optimization
problem. Therefore, the proposed algorithm can not only achieve good
achievable rates but also ensure sensing performance.

\subsection{Robustness Evaluation}

Due to the difficulty of perfectly estimating channel parameters and
the position parameters of sensing targets, the network inputs can
contain a certain degree of estimation noise. Hence, it is necessary
to further evaluate whether the designed network can exhibit certain
robustness to estimation noises.

\begin{figure}[htbp]
\begin{centering}
\includegraphics[width=0.38\textwidth]{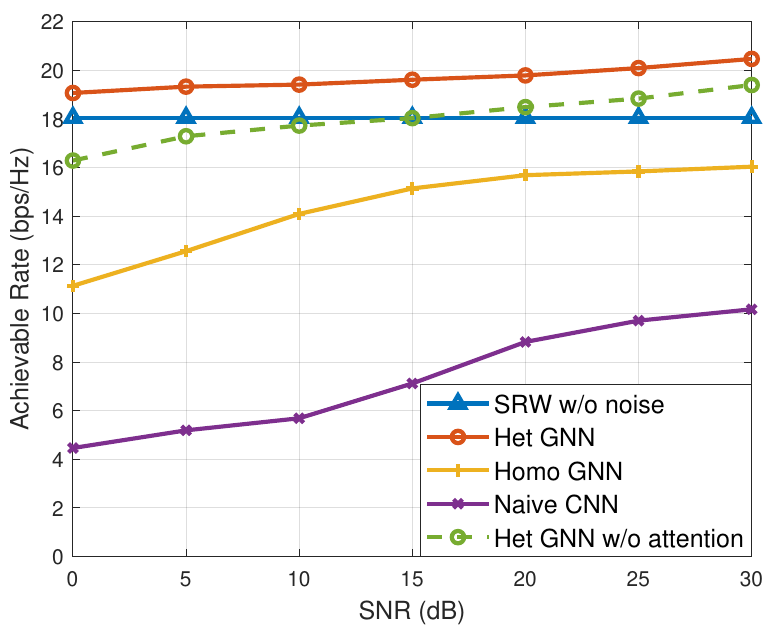} 
\par\end{centering}
\caption{Achievable rate versus SNR of different algorithms.}
\label{img::SNR_ISAC} 
\end{figure}

\begin{figure}[htbp]
\begin{centering}
\includegraphics[width=0.38\textwidth]{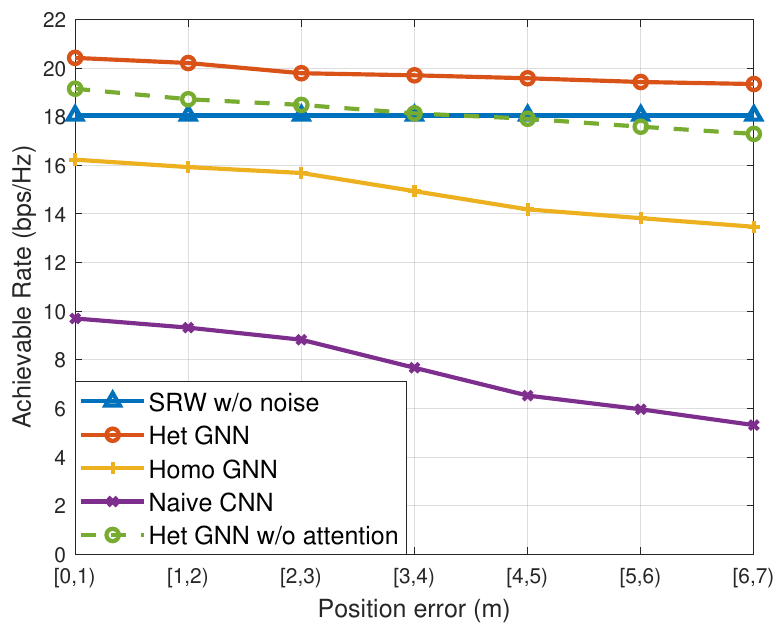} 
\par\end{centering}
\caption{Achievable rate versus position error of different algorithms.}
\label{img::Position_error_ISAC} 
\end{figure}

Let $\hat{\boldsymbol{h}}_{k,n}$ and $\hat{\boldsymbol{a}}_{t}(\hat{\theta}_{n},\hat{\beta}_{n})$
represent the noisy communication node features and sensing node features
respectively. The error of $\hat{\boldsymbol{h}}_{k,n}$ depends on
the accuracy of channel estimation, which incorporates the signal-to-noise
ratio (SNR). The noise in $\hat{\boldsymbol{a}}_{t}(\hat{\theta}_{n},\hat{\beta}_{n})$
comes from position error. Denote the estimated coordinates of the
sensing target by $\hat{\boldsymbol{t}}=\left[\hat{t}_{x},\hat{t}_{y},\hat{t}_{z}\right]^{T}=\left[t_{x}+\Delta t_{x},t_{y}+\Delta t_{y},t_{z}+\Delta t_{z}\right]^{T}$,
where $\Delta t_{x}$, $\Delta t_{y}$ and $\Delta t_{z}$ represent
the position estimation errors in different dimensions. With estimated
coordinates $\hat{\boldsymbol{t}}$, the angle estimation values $\hat{\theta}_{n}$
and $\hat{\beta}_{n}$ of the $n$-th BS can be calculated base on
\eqref{eq:theta} and \eqref{eq:beta}. Then we can calculate the
noisy transmit steering vector $\hat{\boldsymbol{a}}_{t}(\hat{\theta}_{n},\hat{\beta}_{n})$
with estimated angles. Therefore, the noise level of the sensing node
feature$\hat{\boldsymbol{a}}_{t}(\hat{\theta}_{n},\hat{\beta}_{n})$
can be measured by position estimation errors $\Delta t_{x}$, $\Delta t_{y}$
and $\Delta t_{z}$. In the simulation process, the input data is
modified to $\{\hat{\mathbf{H}}^{i},\hat{\mathbf{A}}_{t}^{i}(\boldsymbol{\theta,\beta})\}_{i=1}^{I_{T}}$,
while the noise-free dataset $\{\mathbf{H}^{i},\mathbf{A}_{t}^{i}(\boldsymbol{\theta,\beta})\}_{i=1}^{I_{T}}$
is used for loss calculation. Other simulation settings remain the
same. The results are shown in Fig. \ref{img::SNR_ISAC} and Fig.
\ref{img::Position_error_ISAC}. The ``LHGNN w/o attention'' algorithm
adopts the same LHGNN but does not use the attention mechanism, i.e.,
the aggregation process of this network only needs to average the
feature values of all neighboring nodes, and other parameter settings
are consistent with the proposed algorithm.

In Fig. \ref{img::SNR_ISAC}, the impact of different levels of channel
estimation error is evaluated, with the initial level of position
estimation error set to $\Delta t_{x},\Delta t_{y},\Delta t_{z}\in\left[2,3\right)$,
meaning the coordinate estimation error is randomly generated from
2 to 3 meters. The maximum transmission power of each BS is set to
$P=20$dBm. The simulation results show that the proposed algorithm
can exhibit good robustness under different levels of channel estimation
noise, achieving good performance even at SNR $=0$dB. Meanwhile,
the simulation results indicate that introducing the attention mechanism
can further enhance the robustness of LHGNN. The reason is that the
attention mechanism allows the network to automatically learn attention
coefficients during the aggregation process, thereby assigning smaller
weight values to neighboring nodes severely contaminated by noise,
resulting in better model robustness. In Homo GNN algorithm, although
attention mechanisms are also used, the lack of heterogeneity in modeling
makes it more susceptible to data noise and resulting in lower robustness. 

In Fig. \ref{img::Position_error_ISAC}, the impact of different levels
of position error is evaluated, with the initial position SNR to be
SNR$=20$dB, and the transmission power to be $P=20$dBm. The simulation
results show that the proposed algorithm can exhibit good robustness
to different levels of position estimation noise for sensing targets,
achieving good performance even when the coordinate error $\Delta t_{x},\Delta t_{y},\Delta t_{z}\in\left[6,7\right)$.
Similarly, the proposed algorithm demonstrates the best robustness
to position error. Therefore, by modeling network heterogeneity and
incorporating an attention mechanism, the proposed algorithm significantly
enhances the robustness of beamforming in CoISAC systems against the
communication channel errors and sensing information errors.

\section{Conclusions}\label{sec:Conclusions}

In this paper, we investigate multi-BS collaborative localization
and communication in CoISAC systems and introduce the LHGNN for joint
beamforming design. By modeling different types of links as heterogeneous
nodes and their interactions as edges, LHGNN effectively captures
the structural dependencies within CoISAC networks, surpassing traditional
bipartite graph models. To enhance robustness against channel and
position estimation errors, we integrate a graph attention mechanism
that dynamically adjusts node and link importance, mitigating the
impact of inaccurate inputs. Numerical results validate that the proposed
attention-enhanced LHGNN achieves superior communication rates while
maintaining sensing performance under power constraints. Additionally,
the proposed method demonstrates strong resilience to system uncertainties,
highlighting its effectiveness in real-world CoISAC applications.

\appendix{}

\subsection{Derivation of \ref{eq:EFIM}}

\label{subsec:Derivation-of}

The FIM w.r.t. unknown position-related parameters $\mathbf{\boldsymbol{\eta}=[\boldsymbol{\omega}^{T},\boldsymbol{\alpha}^{T},\sigma^{2}]^{T}}$
is

\[
\mathbf{J}_{\boldsymbol{\eta}}=\left[{\begin{array}{ccc}
\mathbf{J}_{\boldsymbol{\omega\omega}} & \mathbf{J}_{\boldsymbol{\omega}\boldsymbol{\alpha}} & 0\\[4pt]
\mathbf{J}_{\boldsymbol{\omega}\boldsymbol{\alpha}}^{T} & \mathbf{J}_{{\boldsymbol{\alpha}\boldsymbol{\alpha}}} & 0\\
0 & 0 & NL/\sigma^{4}
\end{array}}\right],
\]
where $\left[\mathbf{J}_{\boldsymbol{\omega\omega}}\right]_{ij}\triangleq\frac{2}{\sigma^{2}}{\rm Re}\left\{ \frac{\partial\boldsymbol{u}^{H}}{\partial\omega_{i}}\frac{\partial\boldsymbol{u}}{\partial\omega_{j}}\right\} $,
$\boldsymbol{u}=(\mathbf{A\odot B}){\rm \boldsymbol{Ps}}$ and $\mathbf{J}_{\boldsymbol{\omega\alpha}}$,
$\mathbf{J}_{\boldsymbol{\alpha\alpha}}$ are defined similarly. Note
that we omit the sample time $t$ for derivation. Performing EFIM
on angle parameter $\boldsymbol{\omega}$, we have

\begin{equation}
\mathbf{J}_{e}\left({\boldsymbol{\omega};\mathbf{P}}\right)=\mathbf{J}_{\boldsymbol{\omega\omega}}-\mathbf{J}_{\boldsymbol{\omega}\boldsymbol{\alpha}}\mathbf{J}_{{\boldsymbol{\alpha}\boldsymbol{\alpha}}}^{-1}\mathbf{J}_{\boldsymbol{\omega}\boldsymbol{\alpha}}^{T}.\label{eq:52}
\end{equation}
Firstly, we calculate the derivatives of $\boldsymbol{u}$ w.r.t.
$\boldsymbol{\omega}$ and $\boldsymbol{\alpha}$

\begin{align*}
\frac{\partial\boldsymbol{u}}{\partial\omega_{i}}= & \left({\frac{\partial\mathbf{A}}{\partial\boldsymbol{\omega}_{i}}\odot\mathbf{B}}\right)\mathbf{P}\boldsymbol{s},i=1,\ldots2N,\\
\frac{\partial\boldsymbol{u}}{\partial\alpha_{u,n}}= & \left({\mathbf{A}\odot\frac{\partial\mathbf{B}}{\partial\alpha_{u,n}}}\right)\mathbf{P}\boldsymbol{s},
\end{align*}
where $\frac{\partial\boldsymbol{u}}{\partial\omega_{i}}$ and $\frac{\partial\boldsymbol{u}}{\partial\alpha_{u,n}}$
can be rewritten in a compact form

\[
\begin{array}{c}
\frac{\partial\boldsymbol{u}}{\partial{\boldsymbol{\omega}}}=\mathbf{D}\bar{\mathbf{P}}\bar{\mathbf{S}}\in\mathbb{C}^{NL\times2N},\frac{\partial\boldsymbol{u}}{\partial{\boldsymbol{\alpha}}}=\tilde{\mathbf{A}}\ddot{\mathbf{P}}\ddot{\boldsymbol{S}}\in\mathbb{C}^{NL\times N^{2}}.\end{array}
\]
with $\tilde{\mathbf{A}}=(\boldsymbol{1}_{N^{2}}^{T}\otimes\mathbf{A})\odot\frac{\partial\mathbf{B}}{\partial\alpha}$,
$\ddot{\mathbf{P}}=\mathbf{I}_{N^{2}}\otimes\mathbf{P}\in\mathbb{C}^{N^{3}L\times N^{2}(K+1)}$,
$\ddot{\boldsymbol{S}}=\mathbf{I}_{N^{2}}\otimes\mathbf{s}$ and $\bar{\mathbf{S}}=\mathbf{I}_{2N}\otimes\mathbf{s}$.
Then, we obtain the FIM expressions as follows,

\begin{equation}
\begin{aligned}\mathbf{J}_{\boldsymbol{\omega}\boldsymbol{\omega}}= & \frac{2}{\sigma^{2}}\textrm{Re}\left\{ {\bar{\mathbf{S}}^{H}\bar{\mathbf{P}}^{H}\mathbf{D}^{H}\mathbf{D}\bar{\mathbf{P}}\bar{\mathbf{S}}}\right\} ,\\
\mathbf{J}_{\boldsymbol{\omega}\boldsymbol{\alpha}}= & \frac{2}{\sigma^{2}}\textrm{Re}\left\{ {\bar{\mathbf{S}}^{H}\bar{\mathbf{P}}^{H}\mathbf{D}^{H}\tilde{\mathbf{A}}\ddot{\mathbf{P}}\ddot{\boldsymbol{S}}}\right\} ,\\
\mathbf{J}_{{\boldsymbol{\alpha}\boldsymbol{\alpha}}}= & \frac{2}{\sigma^{2}}\textrm{Re}\left\{ {\ddot{\boldsymbol{S}}^{H}\ddot{\mathbf{P}}^{H}\tilde{\mathbf{A}}^{H}\tilde{\mathbf{A}}\ddot{\mathbf{P}}\ddot{\boldsymbol{S}}}\right\} .
\end{aligned}
\label{eq:53}
\end{equation}
Using the fact $\mathbf{XY}(\mathbf{Y}^{H}\mathbf{X}^{H}\mathbf{XY})^{\dagger}\mathbf{Y}^{H}\mathbf{X}^{H}=\mathbf{X}(\mathbf{X}^{H}\mathbf{X})^{\dagger}\mathbf{X}^{H}=\mathbf{\Pi}_{\mathbf{X}}^{\perp}$
where $\mathbf{X}$ and $\mathbf{Y}$ are full row and full column
rank matrix, plugging \eqref{eq:53} into \eqref{eq:52}, we obtain

\begin{equation}
\mathbf{J}_{e}\left({\boldsymbol{\omega};\mathbf{P}}\right)=\frac{2}{\sigma^{2}}\textrm{Re}\left\{ {\bar{\mathbf{S}}^{H}\bar{\mathbf{P}}^{H}\mathbf{D}^{H}{\mathbf{\Pi_{\tilde{A}}^{\perp}}D}\bar{\mathbf{P}}\bar{\mathbf{S}}}\right\} ,
\end{equation}
with

\[
\begin{aligned}\left[{\mathbf{J}_{e}\left({\boldsymbol{\omega};\mathbf{P}}\right)}\right]_{ij}= & \boldsymbol{p}^{c,H}\mathbf{D}_{i}^{H}{\mathbf{\Pi_{\tilde{A}}^{\perp}}D}_{j}\boldsymbol{p}^{c}\sum_{t=1}^{T}s_{c}s_{c}^{\ast}/T\\
 & +\,\sum_{k=1}^{K}\boldsymbol{p}_{k}^{p,H}\mathbf{D}_{i}^{H}{\mathbf{\Pi_{\tilde{A}}^{\perp}}D}_{j}\boldsymbol{p}_{k}^{p}\sum_{t=1}^{T}s_{k}s_{k}^{\ast}/T,\\
= & \frac{2}{\sigma^{2}}\textrm{Re}{\Biggl\{{\left({\boldsymbol{p}^{c,H}\mathbf{D}_{i}^{H}{\mathbf{\Pi_{\tilde{A}}^{\perp}}D}_{j}\boldsymbol{p}^{c}}\right)\odot\boldsymbol{\Lambda}_{c}}}\\
 & +\,{\sum_{k=1}^{K}\left({\bar{\mathbf{P}}_{k}^{H}\mathbf{D}^{H}{\mathbf{\Pi_{\tilde{A}}^{\perp}}D}\bar{\mathbf{P}}_{k}}\right)\odot\boldsymbol{\Lambda}_{k}}\Biggr\},
\end{aligned}
\]
where $t$ is sample time index and $\mathbf{D}_{i}$ is the $i$-th
block of $\mathbf{D}$. Since the covariance matrix of ${\rm \boldsymbol{s}}$
is identity matrix, we can obtain EFIM expression as in \eqref{eq:EFIM}

\bibliographystyle{IEEEtran}
\bibliography{reference.bib}

\begin{thebibliography}{10}
\providecommand{\url}[1]{#1}
\csname url@samestyle\endcsname
\providecommand{\newblock}{\relax}
\providecommand{\bibinfo}[2]{#2}
\providecommand{\BIBentrySTDinterwordspacing}{\spaceskip=0pt\relax}
\providecommand{\BIBentryALTinterwordstretchfactor}{4}
\providecommand{\BIBentryALTinterwordspacing}{\spaceskip=\fontdimen2\font plus
\BIBentryALTinterwordstretchfactor\fontdimen3\font minus \fontdimen4\font\relax}
\providecommand{\BIBforeignlanguage}[2]{{%
\expandafter\ifx\csname l@#1\endcsname\relax
\typeout{** WARNING: IEEEtran.bst: No hyphenation pattern has been}%
\typeout{** loaded for the language `#1'. Using the pattern for}%
\typeout{** the default language instead.}%
\else
\language=\csname l@#1\endcsname
\fi
#2}}
\providecommand{\BIBdecl}{\relax}
\BIBdecl

\bibitem{tan2021integrated}
D.~K.~P. Tan, J.~He, Y.~Li, A.~Bayesteh, Y.~Chen, P.~Zhu, and W.~Tong, ``Integrated sensing and communication in {6G}: Motivations, use cases, requirements, challenges and future directions,'' in \emph{2021 1st IEEE International Online Symposium on Joint Communications \& Sensing (JC\&S)}.\hskip 1em plus 0.5em minus 0.4em\relax IEEE, 2021, pp. 1--6.

\bibitem{wu2015cloudRAN}
J.~Wu, Z.~Zhang, Y.~Hong, and Y.~Wen, ``Cloud radio access network {(C-RAN)}: a primer,'' \emph{IEEE network}, vol.~29, no.~1, pp. 35--41, 2015.

\bibitem{liu2020cloud}
A.~Liu, L.~Lian, V.~Lau, G.~Liu, and M.-J. Zhao, ``Cloud-assisted cooperative localization for vehicle platoons: A turbo approach,'' \emph{IEEE Transactions on Signal Processing}, vol.~68, pp. 605--620, 2020.

\bibitem{liu2022survey}
A.~Liu, Z.~Huang, M.~Li, Y.~Wan, W.~Li, T.~X. Han, C.~Liu, R.~Du, D.~K.~P. Tan, J.~Lu \emph{et~al.}, ``A survey on fundamental limits of integrated sensing and communication,'' \emph{IEEE Communications Surveys \& Tutorials}, vol.~24, no.~2, pp. 994--1034, 2022.

\bibitem{chen2023comp_clustering}
L.~Chen, X.~Qin, Y.~Chen, and N.~Zhao, ``Joint waveform and clustering design for coordinated multi-point {DFRC} systems,'' \emph{IEEE Trans. Commun.}, vol.~71, no.~3, pp. 1323--1335, 2023.

\bibitem{babu2024precoding}
N.~Babu, C.~Masouros, C.~B. Papadias, and Y.~C. Eldar, ``Precoding for multi-cell {ISAC}: from coordinated beamforming to coordinated multipoint and bi-static sensing,'' \emph{arXiv preprint arXiv:2402.18387}, 2024.

\bibitem{gao2023SPEB}
P.~Gao, L.~Lian, and J.~Yu, ``Cooperative {ISAC} with direct localization and rate-splitting multiple access communication: A pareto optimization framework,'' \emph{IEEE Journal on Selected Areas in Communications}, vol.~41, no.~5, pp. 1496--1515, 2023.

\bibitem{ISACrobust:yang2024coordinatedy}
X.~Yang, Z.~Wei, J.~Xu, Y.~Fang, H.~Wu, and Z.~Feng, ``Coordinated transmit beamforming for networked {ISAC} with imperfect {CSI} and time synchronization,'' \emph{IEEE Transactions on Wireless Communications}, 2024.

\bibitem{ISACrobust:zhang2024energy}
H.~Zhang, H.~Sun, T.~He, W.~Xiang, and R.~Q. Hu, ``Energy efficient robust beamforming for vehicular {ISAC} with imperfect channel estimation,'' in \emph{2024 IEEE International Conference on Communications Workshops (ICC Workshops)}.\hskip 1em plus 0.5em minus 0.4em\relax IEEE, 2024, pp. 1864--1869.

\bibitem{ISACLSTM_liu2022learning}
C.~Liu, W.~Yuan, S.~Li, X.~Liu, H.~Li, D.~W.~K. Ng, and Y.~Li, ``Learning-based predictive beamforming for integrated sensing and communication in vehicular networks,'' \emph{IEEE Journal on Selected Areas in Communications}, vol.~40, no.~8, pp. 2317--2334, 2022.

\bibitem{liu2022ppo_bf}
X.~Liu, H.~Zhang, K.~Long, M.~Zhou, Y.~Li, and H.~V. Poor, ``Proximal policy optimization-based transmit beamforming and phase-shift design in an {IRS-aided} {ISAC} system for the thz band,'' \emph{IEEE Journal on Selected Areas in Communications}, vol.~40, no.~7, pp. 2056--2069, 2022.

\bibitem{yang2024deep}
R.~Yang, Z.~Zhu, J.~Zhang, S.~Xu, C.~Li, Y.~Huang, and L.~Yang, ``Deep learning-based joint transmit beamforming for dual-functional radar-communication system,'' \emph{IEEE Transactions on Wireless Communications}, 2024.

\bibitem{qi2024deep_uplink_ISAC}
Q.~Qi, X.~Chen, C.~Zhong, C.~Yuen, and Z.~Zhang, ``Deep learning-based design of uplink integrated sensing and communication,'' \emph{IEEE Transactions on Wireless Communications}, 2024.

\bibitem{ISACCNN_chen2024complex}
X.~Chen, Z.~Feng, J.~A. Zhang, F.~Gao, X.~Yuan, Z.~Yang, and P.~Zhang, ``Complex {CNN} {CSI} enhancer for integrated sensing and communications,'' \emph{IEEE Journal of Selected Topics in Signal Processing}, 2024.

\bibitem{ISACFNN_pulkkinen2024model}
P.~Pulkkinen and V.~Koivunen, ``Model-based online learning for active {ISAC} waveform optimization,'' \emph{IEEE Journal of Selected Topics in Signal Processing}, 2024.

\bibitem{SUV_shi2023largescale}
Y.~Shi, L.~Lian, Y.~Shi, Z.~Wang, Y.~Zhou, L.~Fu, L.~Bai, J.~Zhang, and W.~Zhang, ``Machine learning for large-scale optimization in {6G} wireless networks,'' \emph{IEEE Communications Surveys \& Tutorials}, 2023.

\bibitem{shen2022graph}
Y.~Shen, J.~Zhang, S.~Song, and K.~B. Letaief, ``Graph neural networks for wireless communications: From theory to practice,'' \emph{IEEE Transactions on Wireless Communications}, vol.~22, no.~5, pp. 3554--3569, 2022.

\bibitem{GNNAPP_yan2024cellfree}
X.~Yan, Z.~Wang, Y.~Jia, Z.~Zhang, and Y.~Huang, ``Access point selection and beamforming design for cell-free network: From fractional programming to {GNN},'' \emph{IEEE Transactions on Wireless Communications}, 2024.

\bibitem{GNNAPP_li2024gnn_isac}
X.~Li, M.~Chen, Y.~Hu, Z.~Zhang, D.~Liu, and S.~Mao, ``Jointly optimizing terahertz based sensing and communications in vehicular networks: A dynamic graph neural network approach,'' \emph{IEEE Transactions on Wireless Communications}, 2024.

\bibitem{GNNAPP_huang2024sub}
Z.~Huang, Z.~Wang, and S.~Chen, ``{Sub-6GHz Assisted mmWave Hybrid Beamforming with Heterogeneous Graph Neural Network},'' \emph{IEEE Transactions on Communications}, 2024.

\bibitem{zhao2020beamspace}
H.~Zhao, N.~Zhang, and Y.~Shen, ``Beamspace direct localization for large-scale antenna array systems,'' \emph{IEEE Transactions on Signal Processing}, vol.~68, pp. 3529--3544, 2020.

\bibitem{ben2009robust}
A.~Ben-Tal, A.~Nemirovski, and L.~El~Ghaoui, ``Robust optimization,'' 2009.

\bibitem{shapiro2021lectures}
A.~Shapiro, D.~Dentcheva, and A.~Ruszczynski, \emph{Lectures on stochastic programming: modeling and theory}.\hskip 1em plus 0.5em minus 0.4em\relax SIAM, 2021.

\bibitem{vaswani2017attention}
A.~Vaswani, N.~Shazeer, N.~Parmar, J.~Uszkoreit, L.~Jones, A.~N. Gomez, {\L}.~Kaiser, and I.~Polosukhin, ``Attention is all you need,'' \emph{Advances in neural information processing systems}, vol.~30, 2017.

\bibitem{gill2019practical}
P.~E. Gill, W.~Murray, and M.~H. Wright, \emph{Practical optimization}.\hskip 1em plus 0.5em minus 0.4em\relax SIAM, 2019.

\bibitem{alkhateeb2019deepmimo}
A.~Alkhateeb, ``{DeepMIMO}: A generic deep learning dataset for millimeter wave and massive {MIMO} applications,'' \emph{arXiv preprint arXiv:1902.06435}, 2019.

\end{thebibliography}



\end{document}